\documentclass[12pt]{iopart}

\usepackage{graphicx}
\usepackage{xcolor}

\begin{document}




\title{Dynamics of phase space vortices in Vlasov plasmas with ion scale inhomogeneity : I Constant frequency drive study}


\author{Sanjeev Kumar Pandey$^{[1]}$, Amudon Chingangbam$^{[2,3]}$  and Rajaraman Ganesh$^{[2,3]}$}

\address{$^{1}$Department of Physics, Indian Institute of Technology (IIT) Madras, Chennai 600036, India}
\address{$^{2}$ Institute for Plasma Research (IPR), Bhat, Gandhinagar 382428, India}
\address{$^{3}$ Homi Bhabha National Institute (HBNI), Mumbai, Maharashtra 400094, India}
\ead{sanju23510@gmail.com and ganesh@ipr.res.in}
\vspace{10pt}
\begin{indented}
   \item July 2026
\end{indented}

\begin{abstract}
Formation dynamics and stability starting from various phase space vortex (PSV) or Bernstein-Greene-Kruskal (BGK) structures i.e electron acoustic wave (EAW), Langmuir (LAN) waves is investigated in the presence of a quasi-stationary ion scale (QSIS) inhomogeneity using high resolution Vlasov-Poisson simulations with VPPM-OMP 1.0 solver. In a one dimensional, collisionless, periodic, unmagnetized plasma with kinetic ions and kinetic electrons, we first create a QSIS inhomogeneity using low amplitude electric field drive at ion acoustic (IA) frequency with $k_{eq}=mk_{min}$ [where $m=2$ is the mode number, $k_{min}^{-1}$ corresponds to the longest scale in the system]. While creating QSIS inhomogeneity, we have demonstrated the existence of ion trapped particle instability (ITPI) which saturates as the amplitude of sideband modes become comparable to that of the primary nonlinear mode (quite analogous to the trapped particle instability in large amplitude electron plasma waves). Also, mode transition from $m=2$ to $m=1$ is observed during relaxation period due to the energy cascading process. Finally, an electron acoustic (EA) perturbation of scale $k_{p}=k_{min}$ [$m=1$] is applied on top of the QSIS inhomogeneity to determine its response in the presence of background ion scale inhomogeneity. Some key observations such as formation of transient PSV, wave-wave mode coupling interaction and various frequency generation alongwith comparative investigation with EA perturbation launched in the absence of ion scale inhomogeneity is also reported.

\end{abstract}
%
\vspace{2pc}
\submitto{\PS}
%
\maketitle
%
%
\section{Introduction}
\label{KIKE_Introduction}

Interaction of electrostatic waves with plasmas has been a topic of extensive research for more than a century. In this context one of the significant interest is to study the existence of a class of inﬁnite family of exact stationary solutions for electrostatic, collisionless plasmas known as Bernstein-Greene-Kruskal (BGK) waves which was reported in 1957 \cite{bgk1957}. The seminal BGK paper \cite{bgk1957} had opened a new paradigm on the ways to construct a large class of non-linear states. These modes are spatially inhomogeneous non-linear mode exhibiting a finite self-consistent electric potential and field structures. In a uniform or homogeneous plasma, after some initial Landau damping \cite{landau1946} a large amplitude electrostatic wave oscillates in amplitude, and finally settles down to a BGK mode, as demonstrated theoretically by O'Neil \cite{oneil1965}. These modes are known to be the ﬁnal saturated state of instabilities in numerical simulations which are stabilized by the formation of trapped particle phase space vortex (PSV) structures by the potential well of the ﬁnite amplitude wave. In the past, various studies have been reported which are associated with the stability and instability of the BGK mode \cite{kds1969,kd1970,goldman1970,rosen1972,Schamel_1975, canosa1976,shoucri1978,shoucri1980,schamel1982,koch1983,ghizzo1988,manfredi1997,manfredi2000,brunetti2000,brunner2004,shoucri_2006,brunner2014,shoucri2017,yang2020,Pandey_2021_TPI_1,Pandey_2021_TPI_2}.  Since, the advent of the BGK modes, tremendous amount of work speculating its existence in nature \cite{Temerin_1982,Franz_1998,Mangeney_1999}, in experiments \cite{lynov_1979,Saeki_1979,Danielson_2004} and in numerical simulations \cite{manfredi1997}.

Generally, to construct these non-linear BGK modes in a numerical simulations, one has to increase the amplitude of either density perturbation or electric field perturbation in the system \cite{manfredi1997,raghunathan2013,pallavithesis,Pandey_2021_TPI_1,Pandey_2021_TPI_2}. However, one such class of non-linear BGK mode exists which can be excited with low amplitude perturbations, known as electron acoustic waves (EAW). In 1991, Holloway and Dorning \cite{Holloway_Dorning_1991}, reported a non-linear structure that could manifest even at the low perturbation amplitudes and termed them as EAW, since their dispersion relation is described by $\omega=1.31kv_{th}$ for all values of wave number $k$ [here, $\omega$ is the wave frequency and $v_{th}$ is the thermal velocity of the plasma electrons]. In linear wave theory form, EAWs experience significant Landau damping \cite{landau1946}, since its phase velocity is comparable to the thermal velocity of electrons. Despite this, EAWs are non-linear structures characterized by electrons that are trapped in the wave troughs, suppressing the Landau damping by flattening the electron velocity distribution at wave phase velocity \cite{valentini_2006,valentni_2025}. In experiments, EAWs are reported in non-neutral plasmas \cite{Anderegg_2009,Anderegg_PRL_2009}. In numerical simulations, several authors \cite{valentini_2006,valentni_2025}, have demonstrated that these waves can be excited by small amplitude drivers applied resonantly over several electron trapping periods. Also, several studies have been reported in the past, which are related to the excitation, stability and associated parametric instabilities of these EAWs \cite{valentini_2006,valentni_2025,Rivera_2025}. Recently, several authors have demonstrated the use of external drives with time dependent frequency $\omega(t)$ or chirp to obtain BGK modes in bounded \cite{Breizman_1997,Eremin_2002,Fajans_2003,Friedland_2004,Peinetti_2005} and periodic \cite{pallavi2016,pallavi2017,pallavithesis} systems.

In the context of the above-said studies, most of them were carried out in a uniform or homogeneous plasma equilibrium. Whereas, in realistic scenarios such as laboratories, tokamaks, and astrophysical plasmas, equilibria are inhomogeneous in nature. So, it becomes important to excite these BGK modes or phase space vortex [PSV] structures in the presence of a self-consistent [quasi-stationary ion scale (QSIS)] background inhomogeneity, in order to understand the complete physical picture and phase space dynamics behind stability of these [BGK/PSV] structures and their interplay with the inhomogeneous ion scale background. With the advent of High Performance Computing [HPC], which helped us to set up very large number of phase space grid sizes and enabled us to run long time simulations to investigate subtle effects, more accurately. Since, there are two different numerical ways to excite these [BGK/PSV] structures [i.e electric field drive with constant and time dependent frequencies], our computational efforts are divided into two companion papers i.e Part I and Part II. In Part I, using constant frequency low amplitude driver, we have addressed the excitation, stability and phase space dynamics of the EAWs in the presence of QSIS background inhomogeneity. Mean while, in Part II, we have implemented time dependent frequency drivers or chirp drive to investigate the long time stability and phase space dynamics of various class of PSVs [i.e EAWs, LAN and Shark/Honeycomb like structures].

In the first part of the work [i.e Part-I], in a 1D collisionless, periodic, unmagnetized plasma, using the high resolution Eulerian VPPM-OMP 1.0 solver, we have investigated the dynamical response of electron acoustic wave [EAW] launched in the presence of an quasi-stationary ion scale [QSIS] inhomogeneity of scale $k_{eq}/k_{min}=m=2$ where $k_{min}=0.4$ [$k_{min}^{-1}$ corresponds to the longest scale in the system] and $m=2$ is the mode number. At first, we have created this QSIS inhomogeneity using low amplitude long interval electric field drive at ion acoustic [IA] frequency $\omega_{IA}^{D}=0.020223$. During this creation process, we have observed the usual decay instability like trapped particle instability [TPI] in large amplitude electron plasma waves lead to phase space vortex merging and transition from $m=2$ to $m=1$ mode. We have termed it as ion trapped particle instability [ITPI]. The primary reason behind the appearance of this ITPI is the energy cascading process due to wave-wave mode coupling interactions, which leads to the amplitude equivalence between the sideband mode and the ion scale mode i.e $k_{eq}$ around $T_{D}^{ion}$ termed ion destabilization time. Next, we launched an EAW perturbation in the presence of the QSIS inhomogeneity and found several interesting features such as transient PSV formation during evolution, generation of the frequencies corresponding to EAW, and Langmuir [LAN] modes etc. In addition, for exact parameters, we have also compared the simulation results with the results of EAW launched in the absence of any inhomogeneity and reported key highlights such as difference in the mode coupling signatures, and final state of the BGK waves at the end of the simulation.

This paper is organized as follows: In Sec. \ref{KIKE_Mathematical_model}, we present the Vlasov-Poisson model equations including the equations for QSIS inhomogeneity creation using ion acoustic drive and electron acoustic perturbation drive. Followed by implemented numerical scheme and diagnostics for analysis in Sec. \ref{KIKE_Numerical_Scheme}. In Sec. \ref{KIKE_Simulation_Results}, we present the simulation results for the construction of QSIS inhomogeneity (in Sec. \ref{KIKE_Equillibrium_Construction}) and response of EAW in the presence of QSIS inhomogeneous background (in Sec. \ref{KIKE_EAW_IAW}) and finally we conclude in Sec. \ref{KIKE_Discussion_conclusion}.

\section{Mathematical model}
\label{KIKE_Mathematical_model}

In the kinetic theory framework, evolution of electron plasma waves (EPW) in an unmagnetized, 1D, collision-less plasma system consisting of background inhomogeneities created due to kinetic ions can be modeled using a set of coupled Vlasov-Poisson equations \cite{manfredi1997,raghunathan2013,sanjeev2021,Pandey_2021_TPI_1,Pandey_2021_TPI_2,pandey_2022_KAW, Pandey_2024,Saini2018,pallavithesis,sanjeevthesis},
\begin{equation}
\frac{\partial f_{e}}{\partial t}+v_{e}\frac{\partial f_{e}}{\partial x}-E_{T}\frac{\partial f_{e}}{\partial v_{e}}=0
\label{EQ_1}
\end{equation}

\begin{equation}
\frac{\partial f_{i}}{\partial t}+v_{i}\frac{\partial f_{i}}{\partial x}+ \left[ \frac{E_{T}}{m_{r}} \right] \frac{\partial f_{i}}{\partial v_{i}}=0
\label{EQ_2}
\end{equation}

\begin{equation}
\frac{\partial E_{T}}{\partial x}= \int f_{i}dv_{i} - \int f_{e}dv_{e}
\label{EQ_3}
\end{equation}
where $f_{i}(x,v_{i},t)$ and $f_{e}(x,v_{e},t)$ are the ion and electron distribution  functions respectively, $v_{i}$ and $v_{e}$ are the ion and electron velocities, $m_{r}=M_{i}/M_{e}$ is the mass ratio of ions to electrons and $E_{T}(x,t)$ is the total electric field given as,  
\begin{equation}
E_{T}(x,t)= E_{s}(x,t) + E_{IAW}^{Equil}(x,t) + E_{EAW}^{Pert}(x,t)
\label{EQ_4}
\end{equation}

\begin{equation}
E_{IAW}^{Equil}(x,t)=E^{D}(x,t) \times g(t)
\label{EQ_5}
\end{equation}

\begin{equation}
g(t)=[1+(t-\tau)^{n}/(\Delta \tau)^{n}]^{-1}~~:~~ E^{D}(x,t)=E_{0}^{D}sin(k_{eq}x \pm \omega_{IA}^{D}t)
\label{EQ_6}
\end{equation}

\begin{equation}
\omega_{IA}^{D}(k_{eq}, m_{r}, T_{r}) = k_{eq} \left[ \frac{1}{m_{r}(\gamma_{e}+ \gamma_{i}T_{r})} \right]^{1/2}
\label{EQ_7}
\end{equation}

\begin{equation}
E_{EAW}^{Pert}(x,t)= E_{0}^{P}sin(k_{p}x \pm \omega_{EA}^{P}t)
\label{EQ_8}
\end{equation}
where $E_{s}(x,t)$ is the self-consistent electric field, $E_{IAW}^{Equil}(x,t)$ is the electric field drive for quasi-stationary ion scale (QSIS) inhomogeneity construction, $g(t)$ is a temporal function multiplied with the drive to excite ions adiabatically without disturbing the electrons, $\omega_{IA}^{D}(k_{eq}, m_{r}, T_{r})$ is the ion acoustic driving frequency normalized with the electron scales, $k_{eq}$ is equilibrium inhomogeneity wave number, $T_{r}=T_{i}/T_{e}$ is the temperature ratio of ions to electrons, $E_{EAW}^{Pert}(x,t)$ is the electron acoustic (EA) electric field perturbation, $k_{p}$ is the perturbation wave number and $\omega_{EA}^{P}$ is the electron acoustic perturbation frequency. 

\begin{figure}
\centerline{\includegraphics[scale=0.50]{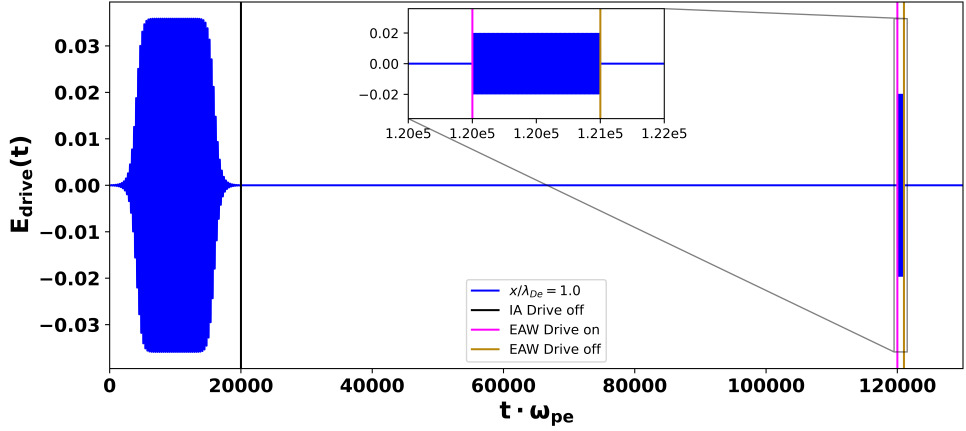}}
\caption{Temporal variation of external electric field drive $E_{drive}(t)$ i.e QSIS inhomogeneity construction drive $[E_{IAW}^{Equil}]$ (defined in Eq. \ref{EQ_5}) applied from $t=0~\omega_{pe}^{-1}$ to $t=20000~\omega_{pe}^{-1}$ and electron acoustic perturbation drive $[E_{EAW}^{Pert}(x,t)]$ (defined in Eq. \ref{EQ_8}). Zoomed inset plot shows the electron acoustic perturbation applied from $t=120000~\omega_{pe}^{-1}$ to $t=121000~\omega_{pe}^{-1}$. Solid lines denotes the various drive on and off times respectively.}
\label{FIG_1_EF_Drive}
\end{figure}

In the above plasma system [Eq. \ref{EQ_1}-\ref{EQ_3}], space is normalized to electron debye length $\lambda_{D_{e}}$, time is normalized to inverse of electron plasma frequency $\omega_{pe}^{-1}$, velocities normalized to electron thermal velocity $v_{the}=\lambda_{De}\omega_{pe}$, electric field normalized to $en_{0}\lambda_{De}/\epsilon_{0}$, and distribution function has been normalized to $n_{0}/\lambda_{De}\omega_{pe}$ where $n_{0}$ is uniform plasma density. The quasi-stationary ion scale (QSIS) inhomogeneity construction $(k_{eq})$ and perturbation $(k_{p})$ length scales are expressed as integer multiples of $k_{min}$ where $k_{min}=2 \pi / L_{max}$. It also enforces periodicity into the system. In this article, we will refer the coupled sideband wave numbers as $k/k_{min}$, whereas rest of the two above mentioned wave numbers as $k_{eq}/k_{min}$ and $k_{p}/k_{min}$ respectively. Fig. \ref{FIG_1_EF_Drive} shows variation of adiabetic QSIS inhomogeneity and abrupt electron acoustic external drives defined by Eq. \ref{EQ_5} and Eq. \ref{EQ_8} which is applied from $t=0~\omega_{pe}^{-1}$ to $t=20000~\omega_{pe}^{-1}$ and $t=120000~\omega_{pe}^{-1}$ to $t=121000~\omega_{pe}^{-1}$ respectively. In the next section, we will brief about the scheme used to solve the set of coupled Vlasov-Poisson equation and parameters used to initiate the simulations. 

\section{Numerical Scheme and Diagnostics}
\label{KIKE_Numerical_Scheme}

Using VPPM-OMP 1.0 an in-house developed OpenMP based Vlasov-Poisson solver, kinetic equations (\ref{EQ_1})-(\ref{EQ_3}) defined in Sec. \ref{KIKE_Mathematical_model} are solved numerically at Institute for Plasma Research Gandhinagar, India. VPPM-OMP 1.0 is an 1D Eulerian solver capable of handling both ion and electron dynamics simultaneously \cite{sanjeev2021,Pandey_2021_TPI_1,Pandey_2021_TPI_2,pandey_2022_KAW,sanjeevthesis}. Piecewise Parabolic Method (PPM) advection scheme proposed by Colella and Woodward \cite{colella1984} along-with time stepping method given by Cheng and Knorr \cite{cheng1976} are implemented in the solver. To solve Poisson equation i.e Eq. \ref{EQ_3}, we have applied a Fourier transform (FT) based method which uses OpenMP enabled FFTW libraries. We set the simulation domain in 1D phase space $(x,v)$ as : $D=[0,L_{max}] \times [-v_{e}^{max},v_{e}^{max}]$, where $L_{max}=2\pi / k_{min}=5\pi$ (as $k_{min}=0.4$) is the system size and $v_{e}^{max}=8.0$ chosen sufficiently large so that electron distribution function (EDF) approaches to zero as $|v|$ asymptotes to $v_{e}^{max}$. Periodic boundary consitions are implemented in both velocity and spatial domains. The simulation domain is discretized into $N_{v}$ grid points in both the ion as well as electron velocity domains and $N_{x}$ grid points in the spatial domain.

Initially at $t=0$, we setup the simulations with normalised Maxwellian distribution function for electrons and ions given as,  
\begin{equation}
f_{e}(x,v_{e},t=0)= \frac{1}{\sqrt{2\pi}} exp \left[ \frac{-v_{e}^{2}}{2} \right]
\label{EQ_9}
\end{equation}

\begin{equation}
f_{i}(x,v_{i},t=0)= \frac{1}{\sqrt{2\pi}}\sqrt{\frac{m_{r}}{T_{r}}} exp \left[ \frac{-m_{r}v_{i}^{2}}{2T_{r}}\right]
\label{EQ_10}
\end{equation}
where $m_{r}$ and $T_{r}$ are the mass ratio and temperature ratio of the ions to electrons. In the present study, to construct the QSIS inhomogeneity profile using kinetic ions self-consistently, we drive the system externally with ion acoustic (IA) electric field $E_{IAW}^{Equil}(x,t)$ given in Eq. \ref{EQ_5}. The function $g(t)$ (Eq. \ref{EQ_6}) helps us to design the drive in such a way that it does not disturb the electron phase space. We apply the IA drive from $t=0~ \omega_{pe}^{-1}$ to $t=20000~ \omega_{pe}^{-1}$ as indicated in Fig. \ref{FIG_1_EF_Drive} with the parameters used in Eq. \ref{EQ_6} are given as : $\tau=10000$, $\delta \tau=6000$, $n=14$, $k_{eq}=0.8$ i.e $k_{eq}/k_{min}=2$, $E_{0}^{D}=0.025$,        $\omega_{IA}^{D}=0.020223$ where $m_{r}=1836$, $T_{r}=0.1$, adiabatic constants $\gamma_{i}=3.0$ and $\gamma_{e}=1.0$ as electrons are isothermal \cite{Chen}. We let the system relax for another $100000~ \omega_{pe}^{-1}$ upto $t=120000~ \omega_{pe}^{-1}$ to get the quasi-stationary ion scale (QSIS) inhomogeneity. After, we apply electron acoustic (EA) drive i.e $E_{EAW}^{Pert}$ (Eq. $\ref{EQ_8}$) from $t=120000~ \omega_{pe}^{-1}$ to $t=121000~ \omega_{pe}^{-1}$ to investigate the response of electron acoustic waves in the presence of QSIS inhomogeneity.    

In the present study, we have used the following numerical diagnostics to analyze the simulation data and interpret the results :
\begin{itemize}

	\item Spatially averaged distribution function, 
	\begin{equation}
	\widehat{f}_{(i,e)}(v,t)= \frac{ \int_{0}^{L_{max}} f_{(i,e)}(x,v,t) dx }{ \int_{ +v_{(i,e)}^{max} }^{ -v_{(i,e)}^{max} } \int_{0}^{L_{max}} f_{(i,e)}(x,v,t) dx dv_{(i,e)} }
	\label{EQ_11}
	\end{equation}
	We plot $\log_{10}[\widehat{f}_{(i,e)}(v,t)]$ versus velocity to inspect about local flattening of the distribution function around resonance locations i.e $v_{(i,e)}^{\phi}=\omega/k$ caused due to wave-particle resonance interactions.

	\item Excess density fraction,
	\begin{equation}
	\frac{\delta n_{(i,e)}}{ n_{0}}(x,t) ~ = ~ \frac{ n_{(i,e)}(x,t)-n_{(i,e)}(x,t=0)}{n_{(i,e)}(x,t=0)}
	\label{EQ_12}
	\end{equation}
	It gives us the information about temporal variation of particle trapping or detrapping fraction at a particular spatial location $x=x_{0}$.
	
	\item Difference of numerical entropy $\Delta S_{(i,e)}(t)$,
	\begin{equation}
	\Delta S_{[i,e]}(t)=S_{[i,e]}(t)-S_{[i,e]}(t=0)
	\label{EQ_13}
	\end{equation} 

	\begin{equation}
	S_{[i,e]}(t)=~-\int_{0}^{L_{max}} \int_{-v_{(i,e)}^{max}}^{+v_{(i,e)}^{max}} f_{[i,e]}(x,v,t) \log f_{[i,e]}(x,v,t)dxdv_{[i,e]}
	\label{EQ_14}
	\end{equation} 
	We have plotted temporal variation of $\Delta S_{[i,e]}$ to show the numerical correctness of the performed simulations. As we know, numerical entropy acts as a measure of `information lost' from the system due to its tendency of monotonic increase of phase space filamentation with time because of the intrinsic property of the Vlasov-Poisson system \cite{manfredi1997,feix}.
		
	\item Total energy of the system,
	\begin{equation}
	TE(t)=KE_{(i,e)}(t)+PE(t)
	\label{EQ_15}
	\end{equation}

	\begin{equation}
	KE_{[i,e]}(t)=\int \int \frac{v_{[i,e]}^{2}}{2}f_{[i,e]}(x,v_{[i,e]},t)dxdv_{[i,e]}
	\label{EQ_16}
	\end{equation}

	\begin{equation}
	PE(t)=\int \frac{1}{2}E^{2}(x,t)dx
	\label{EQ_17}
	\end{equation}
	where $KE_{[i,e]}(t)$ is the kinetic and $PE(t)$ is the potential energy of the system. We plot the relative difference of kinetic energy $\Delta KE_{[i,e]}(t)= KE_{[i,e]}(t)-KE_{[i,e]}(0)$, potential energy $\Delta PE(t)= PE(t)-PE(0)$ and total energy $\Delta TE(t)= TE(t)-TE(0)$ with respect to time to examine the conservation and long time steady state solutions for a specified grid resolution of $[N_{x} \times N_{v}]$ in both ion and electron phase space.  
	
\end{itemize}

\begin{figure}
\centerline{\includegraphics[scale=0.55]{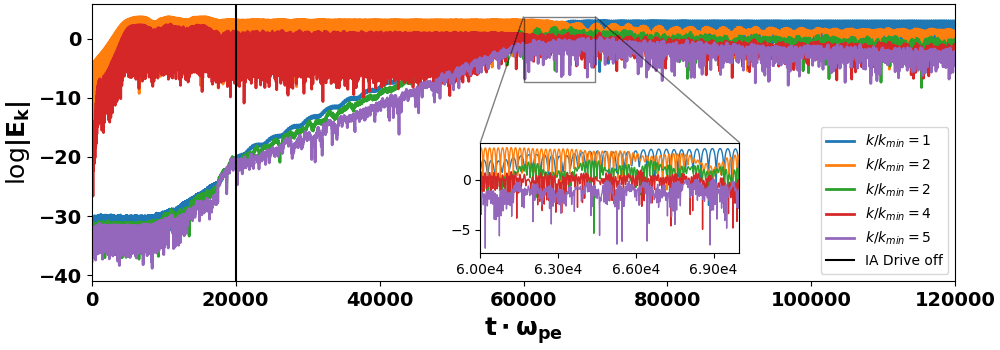}}
\caption{Temporal evolution of QSIS inhomogeneity mode i.e $k_{eq}/k_{min}=2$ and coupled sideband modes i.e $k/k_{min}=1,~3,~4,~5$ driven adiabatically using IA drive [Eq. \ref{EQ_5}] with driving frequency $\omega_{IA}^{D}=0.020223$ upto $t=20000~\omega_{pe}^{-1}$ indicated by the solid line. Inset plot shows zoomed variation from $t=60000~\omega_{pe}^{-1}$ to $t=70000~\omega_{pe}^{-1}$ indicating the amplitude equivalence between equilibrium and sideband modes.}
\label{1_LNEK_EQC}
\end{figure}

\section{Simulation Results}
\label{KIKE_Simulation_Results}
In first part of this Section, we present the simulation results on how the QSIS inhomogeneity is constructed using adiabatic IA electric field drive and in the second part, the response of an EA perturbation launched in the background of the created QSIS inhomogeneity. Also, we compare these results with the EA wave launched in the homogeneous plasma i.e in the absence of any background inhomogeneity. 

\subsection{Construction of Inhomogeneity Using Kinetic Ions}
\label{KIKE_Equillibrium_Construction}
As discussed briefly in Sec. \ref{KIKE_Mathematical_model} and Sec. \ref{KIKE_Numerical_Scheme}, numerical simulations were carried out with kinetic ions and kinetic electrons. In order to create background QSIS inhomogeneity self-consistently, we initiate the simulation parameters as discussed in Sec. \ref{KIKE_Numerical_Scheme}. Both ion and electron phase space grid discretizations were set to $[N_{x} \times N_{v}] = [1024 \times 6000]$. Evolution of Fourier mode amplitude of electric field for mode $k$ i.e $E_{k}(t)$ is shown in Fig. \ref{1_LNEK_EQC}, which is obtained as,
    \begin{equation}
	E(x,t) = \sum_{k} E_{k}(t) e^{-ikx}
    \label{EQ_18}
    \end{equation}
where $E(x,t)$ is the total electric field obtained by solving Poisson's equation [Eq. \ref{EQ_3}]. Fig. \ref{1_LNEK_EQC} shows the time evolution of QSIS inhomogeneity mode i.e $k_{eq}/k_{min}=2$ where $k_{min}=0.4$ and coupled interacting sideband modes i.e $k/k_{min}=1,~3,~4,~5$, which are driven adiabatically using IA drive defined in Eq. \ref{EQ_5} upto $t<=20000~\omega_{pe}^{-1}$ with IA driving frequency $\omega_{IA}^{D}=0.020223$. In the previous studies, several authors \cite{kaw1973,sanjeev2021,Pandey_2021_TPI_1,Pandey_2021_TPI_2} have demonstrated that due to the non-linear perturbation amplitude in multi-mode inhomogeneous plasma system, formation of coupled sideband modes takes place with with wave numbers $|k \pm Nk_{0}|$, where $N$ is an integer refereed as coupling parameter which depends on  the inhomogeneity amplitude as well as scale [i.e $N \sim A/\gamma k_{0}^{2}$ \cite{kaw1973}, in long wavelength limit $k \longrightarrow 0$], $k$ and $k_{0}$ are the perturbation and inhomogeneity scales respectively.    

\begin{figure}
\centerline{\includegraphics[scale=0.40]{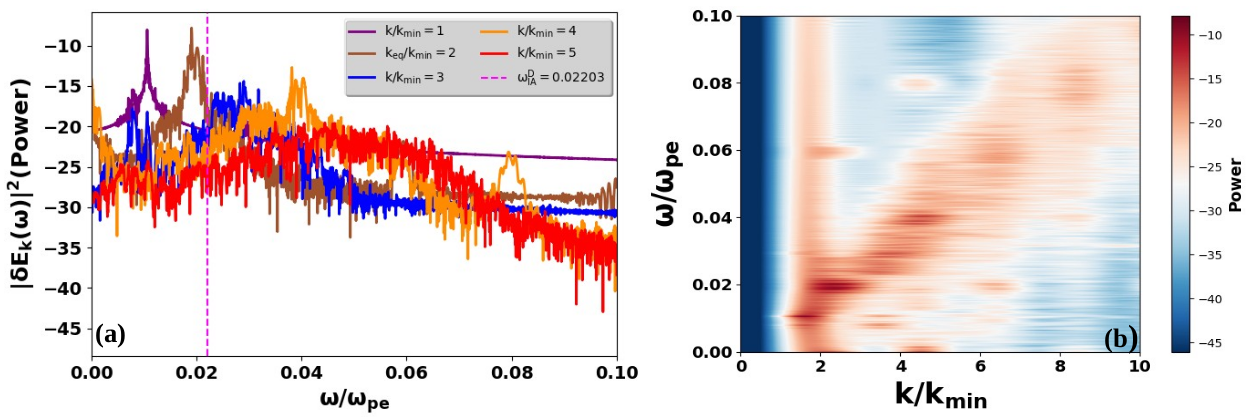}}
\caption{Variation of (a) $|\delta E_{k}(\omega)|^{2}$ vs oscillation frequency $\omega/\omega_{pe}$ using 1D fast Fourier transform (FFT) analysis for QSIS inhomogeneity $k_{eq}/k_{min}=2$ and sideband modes $k/k_{min}=1,3,4,5$ and (b) 2D $(\omega,k)$ power spectrum for adiabatic equilibrium IA drive [Eq. \ref{EQ_5}] case with $k_{min}=0.4$, $E_{0}^{D}=0.025$ and $\omega_{IA}^{D}=0.020223$. Oscillation frequency corresponding to the maximum amplitude for each interacting mode in (a) are listed in Table. \ref{TABLE_1}. The vertical line in (a) is included for IA driving frequency $\omega_{IA}^{D}=0.020223$.  From (b) one can infer that the major power is deposited around QSIS inhomogeneity $k_{eq}/k_{min}$ mode. Also, mode coupling leading to distribution of power can be seen between QSIS inhomogeneity and sideband modes $k/k_{min}=1,3,4,5$.}
\label{1_1DFFT_2DPS_EQC}
\end{figure}

In the present work, we consider $N \sim 3$, since we can not directly determine it due to inherent approximations and have plotted coupled interacting sideband modes $k/k_{min}=1,~3,~4,~5$ as shown in Fig. \ref{1_LNEK_EQC}. Due to the mode coupling caused by non-linearity, a group of adjacent sideband modes are generated and starts to grow from base amplitude of $10^{-30}$ order to almost close to the order of QSIS inhomogeneity mode $k_{eq}/k_{min}=2$. A plausible reason behind the growth of these sideband modes i.e existence of energy exchange phenomenon by wave-wave interactions. As the Fourier mode amplitude of the sideband modes becomes comparable to QSIS inhomogeneity mode around time $T_{D}^{ion}$ i.e $|E_{k_{eq}}(T_{D}^{ion})| \sim E_{k_{eq} \pm Nk}(T_{D}^{ion})|$ due to sideband growth, we observe destabilization effect leading to self-detrapping of particles in ion phase space. Fig. \ref{1_LNEK_EQC} inset plot shows zoomed variation from $t=60000~\omega_{pe}^{-1}$ to $t=70000~\omega_{pe}^{-1}$ indicating the amplitude equivalence between equilibrium and sideband modes. Also, the detrapping time $T_{D}^{ion}$ is approximately equal to $65000~\omega_{pe}^{-1}$. Similar destabilization effect in the electron phase space was observed by the Authors \cite{Pandey_2021_TPI_1,Pandey_2021_TPI_2} in large amplitude Langmuir perturbation cases with immobile ions and kinetic electron Vlasov plasma system. 

\begin{table}
\caption[Mode oscillation frequency and phase velocity corresponding to the QSIS inhomogeneity and interacting sideband modes ]{Mode frequency $(\omega_{k})$ and phase velocity $(v_{\phi,k}^{i}=\omega_{k}/k)$ corresponding to the QSIS inhomogeneity and interacting sideband modes calculated using 1D FFT analysis for IA drive case with $k_{eq}/k_{min}=2$, $k_{min}=0.4$ $E_{0}^{D}=0.025$, $\omega_{IA}^{D}=0.020223$, $\Delta t_{IA}=20000~\omega_{pe}$ $m_{r}=1836$ and $T_{r}=0.1$.}   
\centering                         
\begin{tabular}{c c c}           
\hline\hline                        
Mode No. & Frequency $[\omega_{k}]$ & Phase velocity $[v_{\phi,k}^{i}=\omega_{k}/k]$  \\ [1.0ex]    
\hline          
$k/k_{min}=1$ & 0.0106 & 0.0265 \\          
$k_{eq}/k_{min}=2$ & 0.0202 & 0.0253   \\
$k/k_{min}=3$ & 0.0290 & 0.0242 \\
$k/k_{min}=4$ & 0.0382 & 0.0239 \\
$k/k_{min}=5$ & 0.0425 & 0.0213 \\ [1ex]
\hline\hline                               
\end{tabular}
\label{TABLE_1}
\end{table}

Using 1D fast Fourier transform (FFT) analysis, in Table. \ref{TABLE_1} we have listed up the oscillation frequency $(\omega_{k})$ and the corresponding ion phase velocities $(v_{\phi,k}^{i}=\omega_{k}/k)$ of all the interacting sideband modes. It is remarkable to notice from Table. \ref{TABLE_1} that all the phase velocities values of equilibrium IA as well as coupled sideband modes fall in the range of 0.0265 to 0.0213. Fig. \ref{1_1DFFT_2DPS_EQC} illustrates Variation of (a) $|\delta E_{k}(\omega)|^{2}$ vs oscillation frequency $\omega/\omega_{pe}$ using 1D fast Fourier transform (FFT) analysis for QSIS inhomogeneity $k_{eq}/k_{min}=2$ and sideband modes $k/k_{min}=1,3,4,5$ and (b) 2D $(\omega,k)$ power spectrum for adiabatic IA drive [Eq. \ref{EQ_5}] case with $k_{min}=0.4$, $E_{0}^{D}=0.025$ and $\omega_{IA}^{D}=0.020223$. Oscillation frequency corresponding to the maximum amplitude for each interacting mode in (a) are listed in Table. \ref{TABLE_1}. From (b) one can infer that the major power is deposited around QSIS inhomogeneity $k_{eq}/k_{min}$ mode. Also, due to non-linear IA drive and energy exchange via mode coupling phenomenon leads to the distribution of power which can be seen across QSIS inhomogeneity and sideband modes $k/k_{min}=1,3,4,5$ respectively.

 In kinetic theory, location of the phase velocity $v_{\phi}=\omega/k$ of a wave is an important quantity since it controls the site of resonance energy exchange interaction occurring between wave and particles. In other words, damping or growth of a wave is decided by the slope of the local velocity distribution function i.e $|\partial f / \partial v|_{v_{\phi}}$ at velocity $v=v_{\phi}$.   When the slope $|\partial f / \partial v|_{v_{\phi}} > 0$, the population of slow moving plasma particles with respect to the wave are less compared to the fast ones which results into energy exchange from particles to wave known as inverse Landau damping. Similarly, when the slope $|\partial f / \partial v|_{v_{\phi}} < 0$, the population of slow moving plasma particles are more compared to the fast particles leading to energy exchange from wave to plasma particles commonly termed as Landau damping \cite{landau1946,Pandey_2021_TPI_1,Pandey_2021_TPI_2}.   

\begin{figure}
\centerline{\includegraphics[scale=0.55]{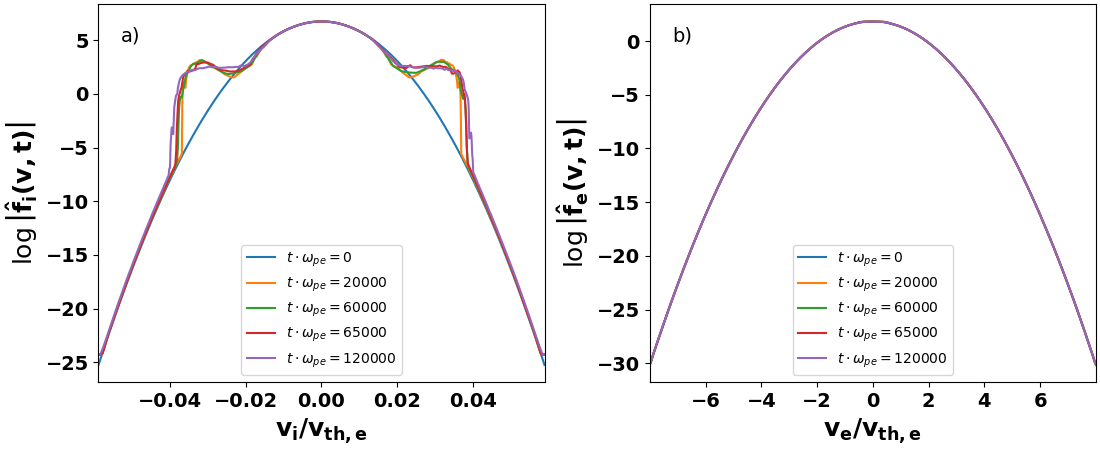}}
\caption{Spatially averaged distribution function $\hat{f}(v,t)$ [Eq. \ref{EQ_11}] plot of (a) ions and (b) electrons at different times i.e $t=0,~20000,~60000,~65000,~120000~\omega_{pe}^{-1}$ and $t=0,~20000,~65000,~120000~\omega_{pe}^{-1}$ respectively for adiabatic IA drive [Eq. \ref{EQ_5}] case. In Fig. (a) one can observe the bump in the ion $\hat{f_{i}}(v,t)$ around phase velocity range of $v_{\phi}=0.0265$ to 0.0213 which corresponds to QSIS inhomogeneity $k_{eq}/k_{min}=2$ and $k/k_{min}=1,3,4,5$ sideband modes listed in Table \ref{TABLE_1}. Meanwhile, in (b) electron $\hat{f_{e}}(v,t)$ remains Maxwellian through out the simulations till $t=120000~\omega_{pe}^{-1}$.}
\label{1_DFE_DFI_EQC}
\end{figure}

Fig. \ref{1_DFE_DFI_EQC} shows the variation of spatially averaged distribution function $\hat{f}_{[i,e]}(v,t)$ [defined by Eq. \ref{EQ_11} in Sec. \ref{KIKE_Numerical_Scheme}] of (a) ions as well as (b) electrons at different times i.e $t=0,~20000,~60000,~65000,~120000~\omega_{pe}^{-1}$ and $t=0,~20000,~65000,~120000~\omega_{pe}^{-1}$ respectively for QSIS inhomogeneity case [defined in Eq. \ref{EQ_5}]. From Fig. \ref{1_DFE_DFI_EQC} (a) we observed that the hump in the ion $\hat{f_{i}}(v,t)$ around phase velocity $v_{\phi}=0.025$ which corresponds to QSIS inhomogeneity $k_{eq}/k_{min}=2$ was initially created due to particle trapping effect of IA drive with $\Delta t_{IA}=20000~\omega_{pe}^{-1}$ resulting into local regions of distribution functions where the slope $|\partial f / \partial v|_{v_{\phi}} > 0$. Afterwards, energy exchange from particles to wave is triggered via inverse Landau damping for sideband modes $k/k_{min}=1,3,4,5$ whose phase velocities $v_{\phi,k}^{i}$ falls into the created 'non-monotonous' structure or hump by IA drive leading to sideband growth as seen in Fig. \ref{1_LNEK_EQC}. Although, we believe that some fraction of energy is also exchanged among interacting modes via wvave-wave mode coupling interactions. Meanwhile, in Fig. \ref{1_DFE_DFI_EQC} (b), spatially averaged electron distribution $\hat{f_{e}}(v,t)$ remains Maxwellian through out the simulation time from $t=0~\omega_{pe}^{-1}$ till $t=120000~\omega_{pe}^{-1}$ indicating that the kinetic ions were absolutely adiabatically driven for a long time period of $t=20000~\omega_{pe}^{-1}$ using IA drive to create the self-consistent inhomogeneity in the ion phase space without any disturbance or structure formation in electron phase space. It is one of the primary reasons that motivated us to attempt such a computationally expensive numerical simulation.  

\begin{figure}
\centerline{\includegraphics[scale=0.65]{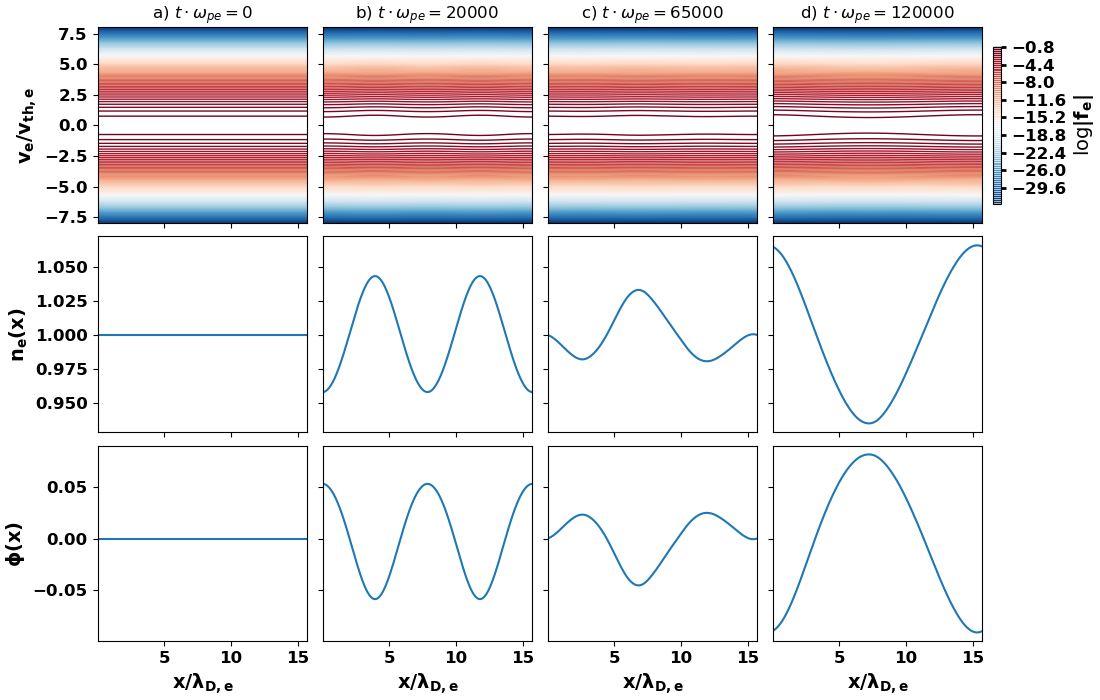}}
\caption{Phase space portrait of electron distribution function $f_{e}(x,v)$ at different times i.e (a) $t=0 ~\omega_{pe}^{-1}$, (b) $t=20000 ~\omega_{pe}^{-1}$, (c) $t=65000 ~\omega_{pe}^{-1}$ and (d) $t=120000 ~\omega_{pe}^{-1}$ during QSIS inhomogeneity creation using external IA drive as shown in Fig. \ref{FIG_1_EF_Drive}. From (a), (b), (c) and (d) one can infer that absence of the phase space structures in the electron phase space signifies the adiabatic nature of the applied IA drive defined in Eq. \ref{EQ_5}.}
\label{1_ELECTRON_CP_EQC}
\end{figure}

\begin{figure}
\centerline{\includegraphics[scale=0.55]{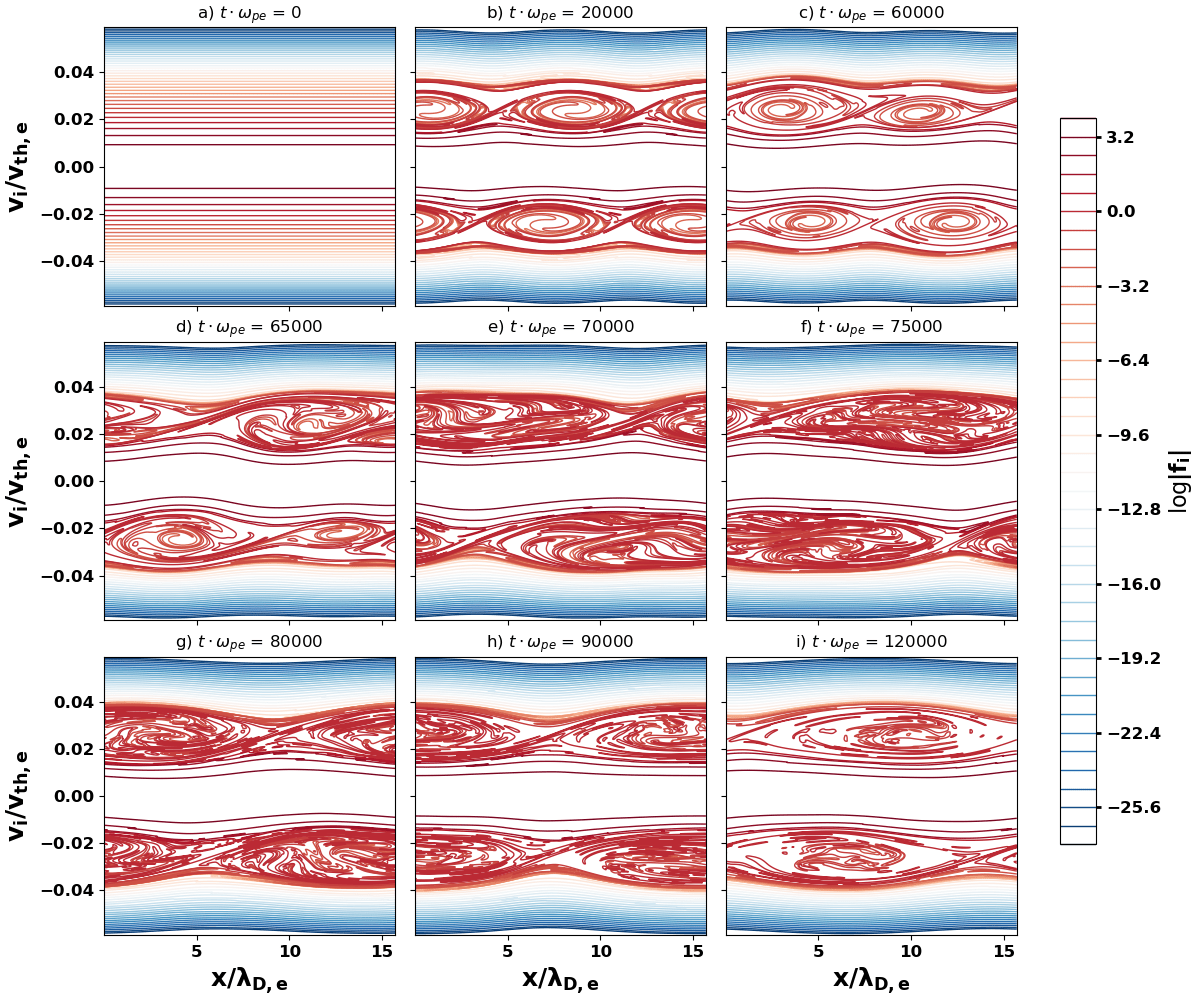}}
\caption{Phase space portrait of ion distribution function $f_{i}(x,v)$ at various times i.e (a) $t= 0~\omega_{pe}^{-1}$, (b) $t= 20000~\omega_{pe}^{-1}$, (c) $t= 60000~\omega_{pe}^{-1}$, (d) $t= 65000~\omega_{pe}^{-1}$, (e) $t= 70000~\omega_{pe}^{-1}$, (f) $t= 75000~\omega_{pe}^{-1}$, (g) $t= 80000~\omega_{pe}^{-1}$, (h) $t= 90000~\omega_{pe}^{-1}$ and (i) $t= 120000~\omega_{pe}^{-1}$ for QSIS construction using adiabatic IA drive [Eq. \ref{EQ_5}]. In Fig. (a), (b) and (c), one can observe the formation of two vortex structures around resonance location $v_{\phi}^{i}=0.02525$ since $ \omega_{IA}^{D}=0.0202$ and $k_{eq}/k_{min}=2$. Destabilization due to trapped particle instability (TPI) in the ion phase space and consequently energy cascading from two vortex structure stream to single vortex structure stream can be seen from Fig. (d) to Fig. (i).}
\label{1_ION_CP_EQC}
\end{figure}

\begin{figure}
\centerline{\includegraphics[scale=0.38]{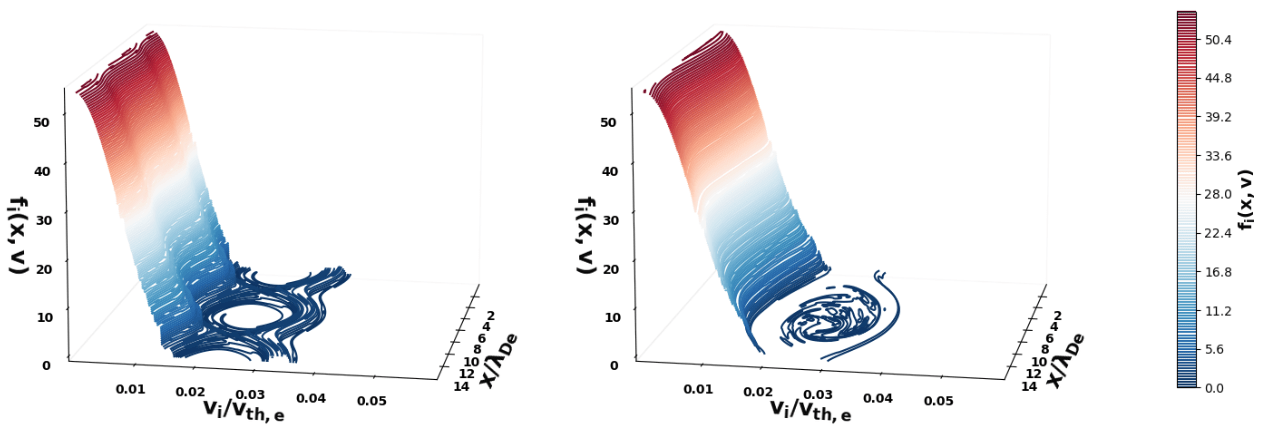}}
\caption{3D surface plot of the ion distribution function $f_{i}(x,v)$ at times $t=20000~\omega_{pe}^{-1}$ (left) and $t=120000~\omega_{pe}^{-1}$ (right) for QSIS inhomogeneity construction run using IA drive [Eq. \ref{EQ_5}] indicating the vortex structure transition from mode $m=2$ to $m=1$ via the energy cascading  between the coupled interacting modes.}
\label{1_3D_ION_SURFACE_PLOT_EQC}
\end{figure}

In Fig. \ref{1_ELECTRON_CP_EQC}, we show the phase space snapshots of electron distribution function $f_{e}(x,v)$ at different times i.e (a) $t=0 ~\omega_{pe}^{-1}$, (b) $t=20000 ~\omega_{pe}^{-1}$, (c) $t=65000 ~\omega_{pe}^{-1}$ and (d) $t=120000 ~\omega_{pe}^{-1}$ evolved during creation of QSIS inhomogeneity profile. From Fig. \ref{1_ELECTRON_CP_EQC} (a), (b), (c) and (d), we observe the absence of the phase space structures in the electron phase space which again establish and signifies the adiabatic nature of the applied IA drive, suggesting that the inhomogeneity created in ion phase space is self-consistent without disturbing the electrons in the system as also pointed by Fig. \ref{1_DFE_DFI_EQC} (b). Similarly,  Fig. \ref{1_ION_CP_EQC} illustrate the evolution of ion phase space at various times i.e (a) $t= 0~\omega_{pe}^{-1}$, (b) $t= 20000~\omega_{pe}^{-1}$, (c) $t= 60000~\omega_{pe}^{-1}$, (d) $t= 65000~\omega_{pe}^{-1}$, (e) $t= 70000~\omega_{pe}^{-1}$, (f) $t= 75000~\omega_{pe}^{-1}$, (g) $t= 80000~\omega_{pe}^{-1}$, (h) $t= 90000~\omega_{pe}^{-1}$ and (i) $t= 120000~\omega_{pe}^{-1}$ for QSIS inhomogeneity construction run using adiabatic IA drive defined by Eq. \ref{EQ_5} with $k_{min}=0.4,~E_{0}^{D}=0.025,~\omega_{IA}^{D}=0.020223$. In Fig. \ref{1_ION_CP_EQC} [(a), (b) and (c)], we have observed that the formation of two vortex structures takes place around resonance ion phase velocity location i.e $v_{\phi}^{i}=0.02525$ since $\omega_{IA}^{D}=0.0202$ and $k_{eq}/k_{min}=2$ as tabulated in Table. \ref{TABLE_1}. Consequently, growth of the generated coupled sideband modes [See Fig. \ref{1_LNEK_EQC}] due to both wave-wave and wave-particle energy exchange interactions takes place which causes amplitude equivalence of QSIS inhomogeneity and sideband mode electric fields that results into the vortex structure destabilization phenomenon at $T_{D}^{ion}=65000~\omega_{pe}^{-1}$ termed as ion trapped particle instability (ITPI) in the ion phase space. Also, as a result of ITPI, energy cascading transition from two vortex structure stream $m=2$ to single vortex structure stream $m=1$ can be seen from Fig. \ref{1_ION_CP_EQC} (d) to (i). The same transition throughout the simulation can be viewed In Fig. \ref{1_3D_ION_SURFACE_PLOT_EQC} where we show 3D surface plot of the ion distribution function $f_{i}(x,v)$ at times $t=20000~\omega_{pe}^{-1}$ (left) and $t=120000~\omega_{pe}^{-1}$ (right).

\begin{figure}
\centerline{\includegraphics[scale=0.55]{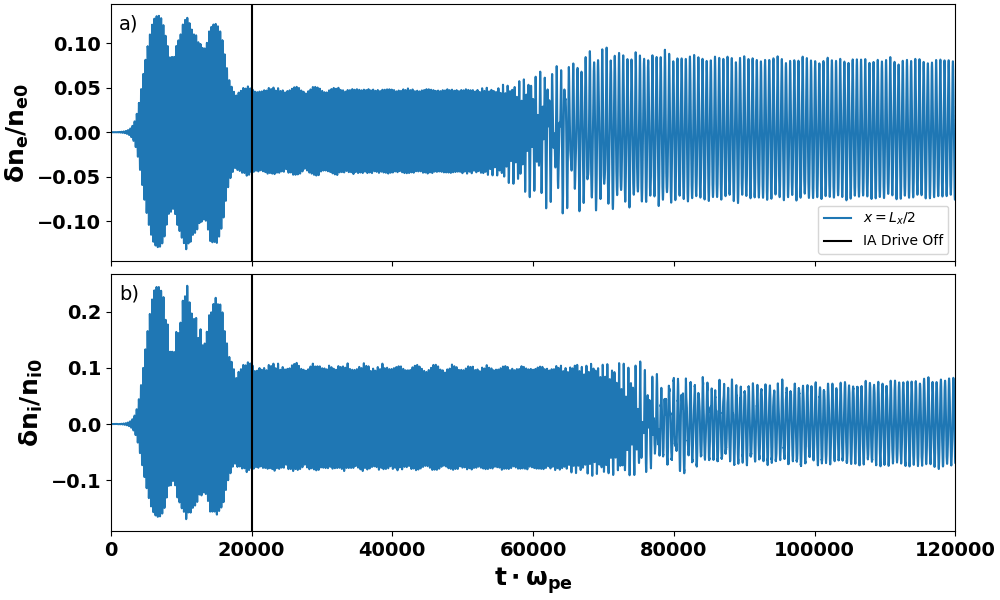}}
\caption{Temporal evolution of the excess density fraction (EDF) defined in Eq. \ref{EQ_12} for (a) electrons and (b) ions with IA drive [\ref{EQ_5}] applied upto $t=20000~\omega_{pe}^{-1}$. Solid black line denotes the time when the IA drve is switched off. One can observe the increase in the trapping fraction of ions till $t=20000~\omega_{pe}^{-1}$ indicating the formation of the vortex structure in ion phase space (as seen in Fig. \ref{1_ION_CP_EQC}). We can also observe the change in the ion density fraction around $T_{D}^{ion}=65000~\omega_{pe}^{-1}$ due to particle detrapping phenomenon caused by the ion trapped particle instability.}
\label{1_EDF_EQC}
\end{figure}

Fig. \ref{1_EDF_EQC} shows temporal evolution of the excess density fraction (EDF) defined by Eq. \ref{EQ_12} in Sec. \ref{KIKE_Numerical_Scheme} for (a) electrons and (b) ions with IA drive applied $0<t<20000~\omega_{pe}^{-1}$. Solid black line denotes the time when the IA drve is switched off. One can observe the increase in the trapping fraction of ions till $t<=20000~\omega_{pe}^{-1}$ indicating the formation of the vortex structure due to particle trapping phenomenon in ion phase space [as seen in Fig. \ref{1_ION_CP_EQC}]. Since, the electric field drive was applied at IA frequency $\omega_{IA}^{D}=0.020223$, we observe more trapping fraction in ion EDF compared to the electron EDF. It is important to note the reduction in the ion density fraction around ion destabilization time $T_{D}^{ion}=65000~\omega_{pe}^{-1}$ indicative of the particle detrapping phenomenon and $m=2$ to $m=1$ mode transition caused by ITPI. Also, the ion excess density fraction attains a finite non-zero $[\partial n_{i}/ n_{0} \neq 0]$ constant value between 5-10 $\%$ at late times till the end of the simulation which implies sustainability of the formed vortex structure.

\begin{figure}
\centerline{\includegraphics[scale=0.50]{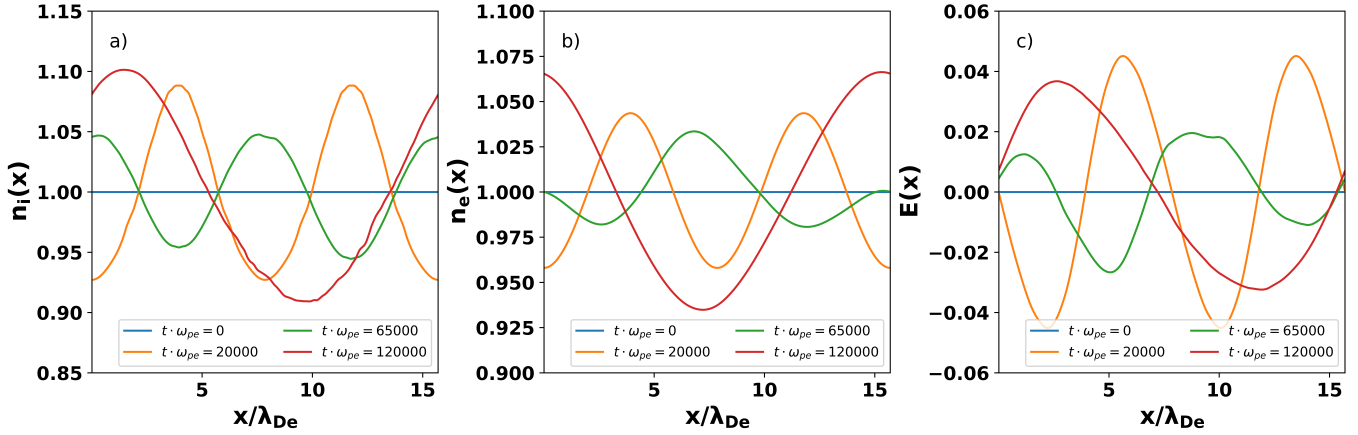}}
\caption{Spatial variation of (a) ion density $n_{i}(x)$, (b) electron density $n_{e}(x)$ and (c) electric field $E(x)$ at different times i.e $t=0,~20000,~65000$ and $120000~\omega_{pe}^{-1}$ for QSIS inhomogeneity construction run with IA drive applied from $t=0~\omega_{pe}^{-1}$ to $t=20000~\omega_{pe}^{-1}$. Comparing Fig. (a) and (b), one can infer that the modulations in the ion density is more compared to the electron density at late times. Fig. (c) indicates that the amplitude of saturated electric field after reaching the steady state around $t=120000~\omega_{pe}^{-1}$ is equal to 0.04.}
\label{1_NI_NE_E_VS_X_EQC}
\end{figure}

In Fig. \ref{1_NI_NE_E_VS_X_EQC}, we present spatial variation of (a) ion density $n_{i}(x)$, (b) electron density $n_{e}(x)$ and (c) electric field $E(x)$ at different times i.e $t=0,~20000,~65000$ and $120000~\omega_{pe}^{-1}$ for QSIS inhomogeneity construction run with IA drive applied between $0<t<20000~\omega_{pe}^{-1}$. On comparison of  Fig. \ref{1_NI_NE_E_VS_X_EQC} (a) and (b), we can say that the density modulations in the ion is more compared to the electron density at late times implying greater particle trapping and thus, structure formation in the corresponding phase space. Fig. \ref{1_NI_NE_E_VS_X_EQC} (c) indicates that the amplitude of saturated electric field after reaching the steady state solution around $t=120000~\omega_{pe}^{-1}$ is equal to 0.04. This value is treated as the reference value in order to decide the perturbation amplitude of the electron acoustic wave (EAW) launched on top of the created QSIS inhomogeneity whose results are presented in the next section i.e Sec. \ref{KIKE_EAW_IAW}.

\begin{figure}
\centerline{\includegraphics[scale=0.65]{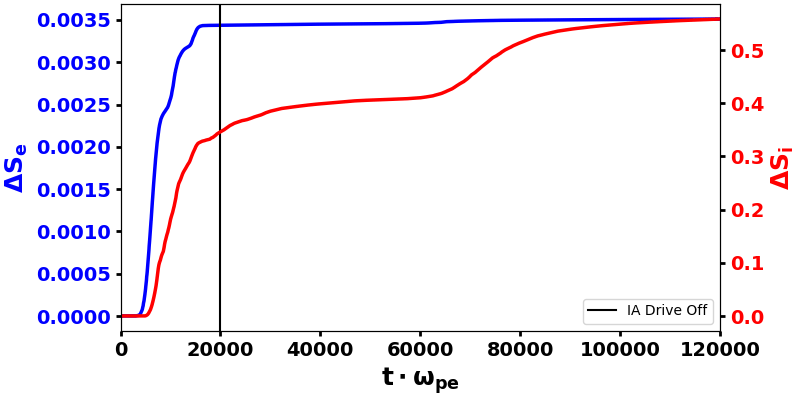}}
\caption{Signature of the difference in entropy $\Delta S$ (defined in Eq. \ref{EQ_13}, \ref{EQ_14}) of ions and electrons with respect to time for QSIS inhomogeneity construction run with IA drive applied from $t=0~\omega_{pe}^{-1}$ to $t=20000~\omega_{pe}^{-1}$ with grid sizes $[N_{x} \times N_{v}=1024 \times 6000]$. Solid line at $t=20000~\omega_{pe}^{-1}$ denotes the IA electric field switch off time. One can notice the order of difference in the numerical entropies of ion and electrons i.e $10^{-1}$ and $10^{-3}$ respectively.}
\label{1_ENTROPY_I_E_EQC}
\end{figure}

Fig. \ref{1_ENTROPY_I_E_EQC} shows the temporal variation of difference in ion and electron entropies i.e  $\Delta S_{i}$ and $\Delta S_{e}$ respectively [defined in Eq. \ref{EQ_13}, \ref{EQ_14} of Sec. \ref{KIKE_Numerical_Scheme}] for QSIS inhomogeneity construction run with IA drive applied from $t=0~\omega_{pe}^{-1}$ to $t=20000~\omega_{pe}^{-1}$ with $\omega_{IA}^{D}=0.020223$. The grid discretization for the run was set to $[N_{x} \times N_{v} =1024 \times 6000]$ in both ion and electron phase space. Solid line at $t=20000~\omega_{pe}^{-1}$ indicates the time when the IA electric field drive was switched off. Generally, evolving distribution function exhibits filamentation phenomenon which tends to generate small scale structure in phase space. When the filamentation reaches the phase space grid size $[N_{x} \times N_{v}]$, the small scale structures generated are dissipated leading to the saturation of numerical entropy with respect to time and a numerically steady state solution. From Fig. \ref{1_ENTROPY_I_E_EQC}, we have observed that $\Delta S_{e}$ increases during the IA drive and saturates immediately after the IA drive is switched off, whereas ion entropy $\Delta S _{i}$ increases monotonically till $t=20000~\omega_{pe}^{-1}$, it saturates for certain time period, then again start to increase around destabilization time $T_{D}^{ion}=65000~\omega_{pe}^{-1}$ and further attains a constant value after $t=110000~\omega_{pe}^{-1}$. It indicates that the chosen grid sizes are sufficient enough to reduce the information loss from the system during the ion trapped particle instability transition taking place in ion phase space resulting to a long time high quality numerical simulation. It also signifies that the inhomogeneity profile obtained in ion phase space at the end of simulation time $t=120000~\omega_{pe}^{-1}$ is a steady state equilibrium solution. Also, We have noticed a remarkable difference in the order of numerical entropies of ion and electrons which are $10^{-1}$ and $10^{-3}$ respectively.

\begin{figure}
\centerline{\includegraphics[scale=0.55]{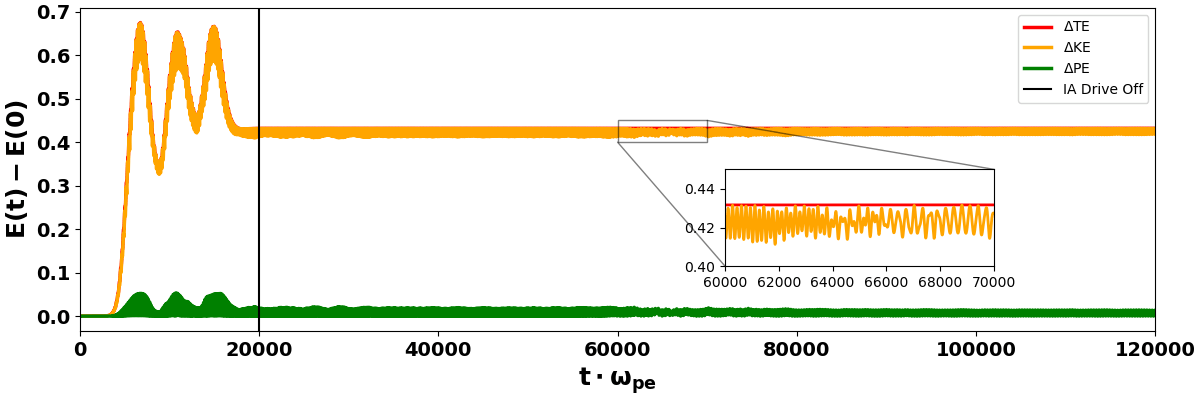}}
\caption{Relative total, kinetic and potential energies $(\Delta TE, \Delta KE, \Delta PE)$ [defined in Eq. \ref{EQ_15}, \ref{EQ_16} and \ref{EQ_17}] signatures with respect to time for QSIS inhomogeneity construction run with IA drive applied from $t=0~\omega_{pe}^{-1}$ to $t=20000~\omega_{pe}^{-1}$. Spatial and velocity $(x,v)$ domain grid discretization for both electrons and ions were set to $[N_{x} \times N_{v}=1024 \times 6000]$. Solid line at $t=20000~\omega_{pe}^{-1}$ denotes the IA electric field switch off time. Inset plot shows the zoomed relative energy variations from $t=60000~\omega_{pe}^{-1}$ to $t=70000~\omega_{pe}^{-1}$. }
\label{1_EC_EQC}
\end{figure}

Fig. \ref{1_EC_EQC} shows the relative total, kinetic and potential energies $(\Delta TE, \Delta KE_{[i,e]}, \Delta PE)$ signatures which are defined by Eq. \ref{EQ_15}, \ref{EQ_16} and \ref{EQ_17} [Sec. \ref{KIKE_Numerical_Scheme}], with respect to time for QSIS inhomogeneity construction run where IA drive applied from $t=0~\omega_{pe}^{-1}$ to $t=20000~\omega_{pe}^{-1}$ with $\omega_{IA}^{D}=0.020223$. Solid line at $t=20000~\omega_{pe}^{-1}$ denotes the IA electric field drive switch off time. From Fig. \ref{1_EC_EQC}, we have observed increase in the $\Delta TE,~ \Delta KE_{[i,e]},~ \Delta PE$ values while the drive was on [$0<t<20000~\omega_{pe}^{-1}$] and then saturation of these signatures with the grid sizes $[N_{x} \times N_{v}=1024 \times 6000]$ when the IA drive is switched off, indicating a very good energy conservation and stable solution. Increase in the $\Delta PE$ value during $0<t<20000~\omega_{pe}^{-1}$ suggest the increase in the particle trapping fraction. Also, in the Fig. \ref{1_EC_EQC}, inset plot shows the zoomed relative energy variations of $\Delta TE, ~ \Delta KE_{[i,e]}$ from $t=60000~\omega_{pe}^{-1}$ to $t=70000~\omega_{pe}^{-1}$ demonstrating constant evolution throughout the ion destabilization time $T_{D}^{ion}=65000~\omega_{pe}^{-1}$ indicative of the absence of any numerical artifact in the obtained solution during ion trapped particle instability (ITPI) transition. In the next section, we will present the simulation results of EAW dynamics launched in the background of the above said QSIS inhomogeneity.    

\subsection{Dynamics of Electron Acoustic Wave (EAW) in the presence of Quasi-Stationary Ion Scale (QSIS) Inhomogeneity}
\label{KIKE_EAW_IAW}

\begin{figure}
\centerline{\includegraphics[scale=0.55]{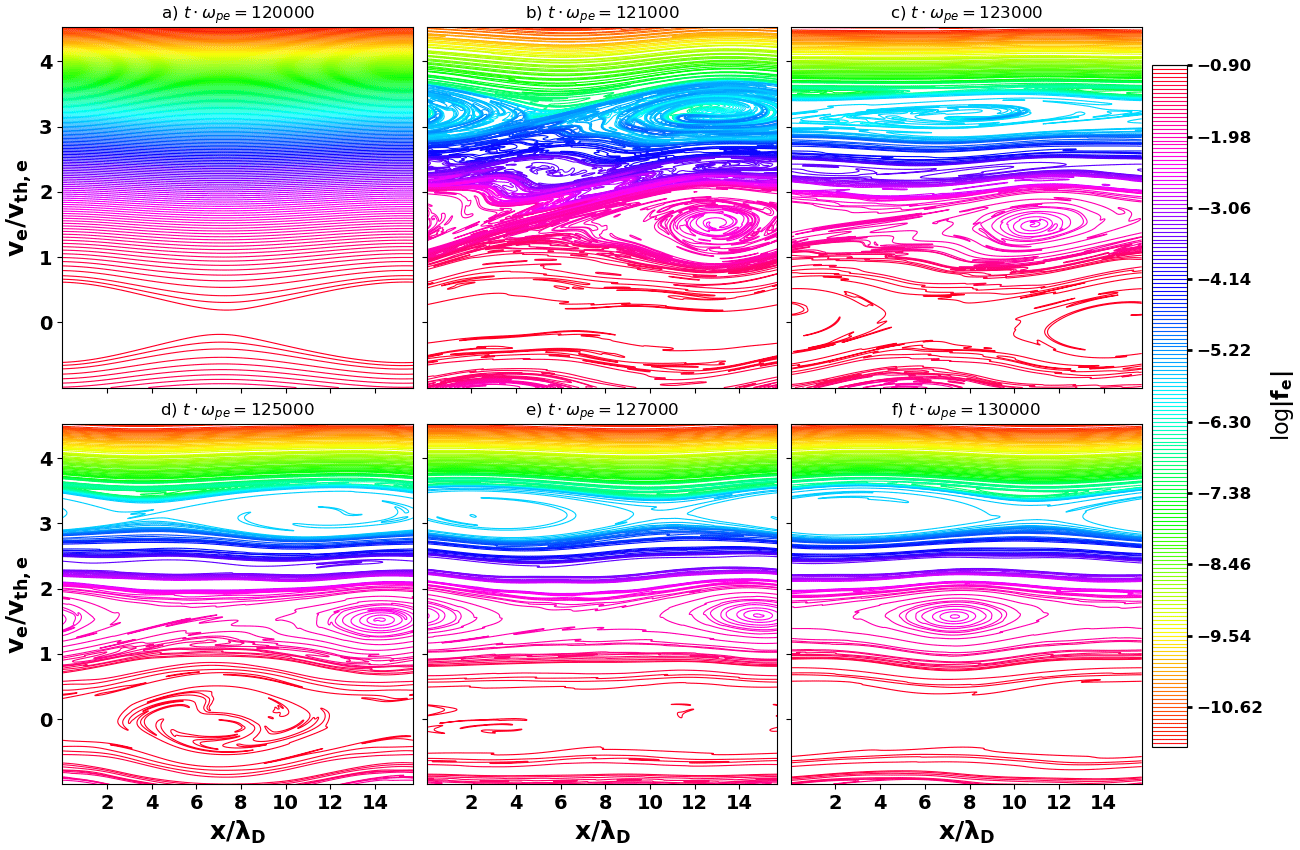}}
\caption{Phase space portrait of electron distribution function $f_{e}(x,v)$ at different times i.e (a) $t=120000~\omega_{pe}^{-1}$, (b) $t=121000 ~\omega_{pe}^{-1}$, (c) $t=123000~\omega_{pe}^{-1}$, (d) $t=125000~\omega_{pe}^{-1}$, (e) $t=127000~\omega_{pe}^{-1}$ and (f) $t=130000~\omega_{pe}^{-1}$ for EAW perturbation driven from $120000~\omega_{pe}^{-1}<t<121000~\omega_{pe}^{-1}$ in the presence of QSIS inhomogeneity case with $k_{p}/k_{min}=1$, $E_{0}^{P}=0.025$ and $\omega_{EA}^{P}=0.624$. From (b) to (f), we can see the formation of EA, Langmuir (LAN) and intermediate structures alongwith their relaxation in the electron phase space due to interaction between perturbation and background ion inhomogeneity modes. Only one side of the phase space is plotted, to bring out the details. }
\label{1_ELECTRON_CP_EAW_PERT}
\end{figure}

\begin{figure}
\centerline{\includegraphics[scale=0.40]{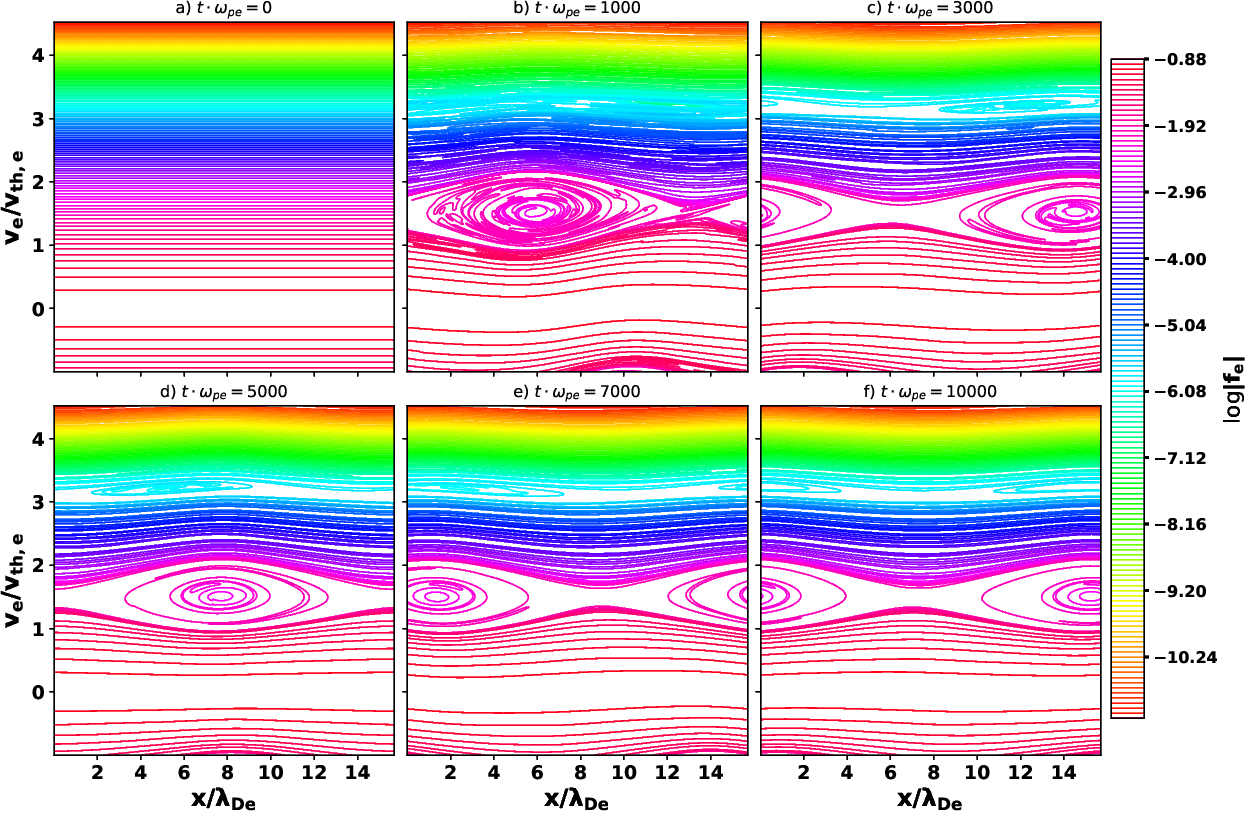}}
\caption{Phase space portrait of electron distribution function $f_{e}(x,v)$ at different times i.e (a) $t=0~\omega_{pe}^{-1}$, (b) $t=1000 ~\omega_{pe}^{-1}$, (c) $t=3000~\omega_{pe}^{-1}$, (d) $t=5000~\omega_{pe}^{-1}$, (e) $t=7000~\omega_{pe}^{-1}$ and (f) $t=10000~\omega_{pe}^{-1}$ for EAW perturbation driven from $0<t<1000~\omega_{pe}^{-1}$ in the absence of QSIS inhomogeneity case with $k_{p}/k_{min}=1$, $E_{0}^{P}=0.025$ and $\omega_{EA}^{P}=0.624$. From (b) to (f), we can see the formation of EA and Langmuir (LAN) in the electron phase space.}
\label{1_CP_ELECTRON_EAW_PERT_HOMO}
\end{figure}

In order to launch an electron acoustic wave (EAW) on top of the created quasi-stationary ion scale (QSIS) inhomogeneity, we initiate the simulations by providing a perturbation of the form $E_{EAW}^{Pert}(x,t)= E_{0}^{P}sin(k_{p}x \pm \omega_{EA}^{P}t)$ [defined by Eq. \ref{EQ_8}] where $E_{0}^{P}=0.025$, $\omega_{EA}^{P}=0.624$, $k_{p}/k_{min}=1:k_{min}=0.4$. The EA perturbation drive is applied between $120000~\omega_{pe}^{-1}<t<121000~\omega_{pe}^{-1}$ and the grid descretization is set to $[N_{x} \times N_{v}=1024 \times 6000]$. In addition, for the comparative study, we have also excited an EA mode with initial driving amplitude $E_{0}=0.025$ and frequency $\omega_{EA}=0.624$ in homogeneous plasma with kinetic electrons and immobile ions. Rest of the simulation parameters are chosen exactly similar to the previous case.   

In the temporal evolution of the Fourier mode $|\delta E_{k}|$ for the EA perturbation in the background of QSIS inhomogeneity, we have observed increase in the amplitude of the EA $|\delta E_{k}|$ Fourier mode i.e $k_{p}/k_{min}=1$ where $k_{min}=0.4$ during $120000~\omega_{pe}^{-1}<t<121000~\omega_{pe}^{-1}$. Generally, in these kind of multi-mode systems, shift in the destabilization time which marks the onset of instability due to trapped particles was reported in large amplitude waves \cite{Pandey_2021_TPI_1,Pandey_2021_TPI_2}. Since, the EA perturbation is applied after the creation of the QSIS inhomogeneity we do not observe any shift in the ITPI destabilization time ($T_{D}^{ion}=65000~\omega_{pe}^{-1}$) mentioned in Sec. \ref{KIKE_Equillibrium_Construction}.

\begin{figure}
\centerline{\includegraphics[scale=0.55]{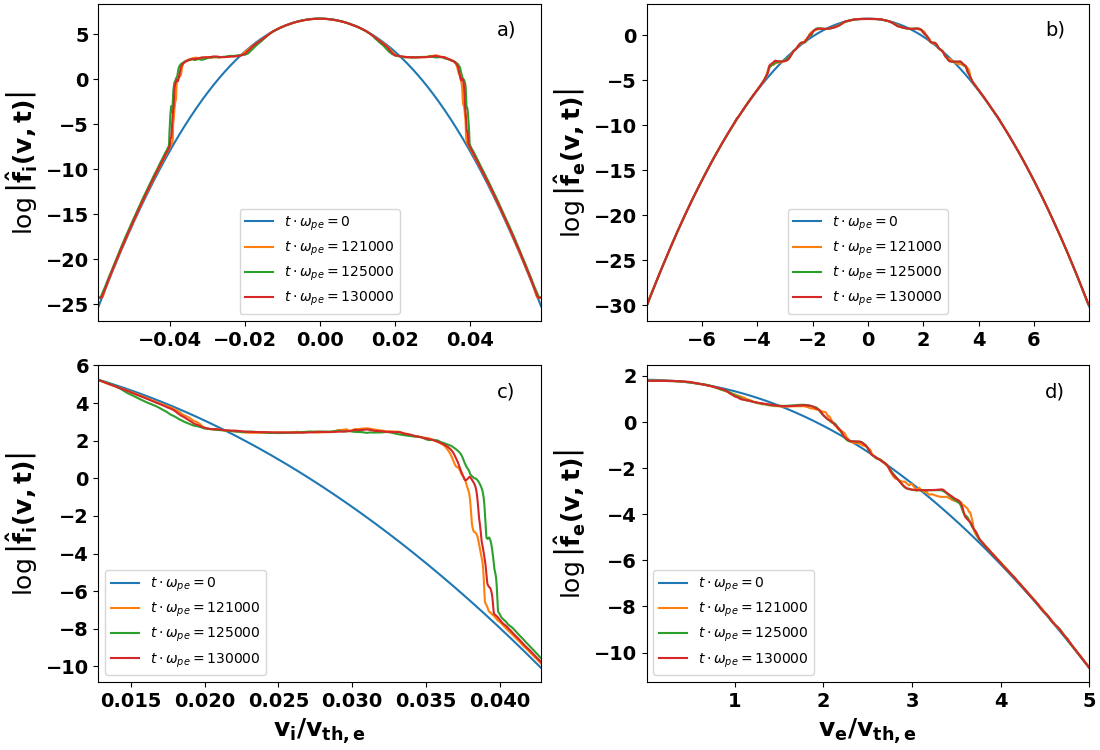}}
\caption{Spatially averaged distribution function $\hat{f}(v,t)$ [Eq. \ref{EQ_11}] plot of (a) ions and (b) electrons at different times i.e $t=121000,~125000,~130000~\omega_{pe}^{-1}$ for EAW perturbation applied from $0<t<121000~\omega_{pe}^{-1}$ in the presence of QSIS inhomogeneity with $k_{p}/k_{min}=1$, $E_{0}^{P}=0.025$ and $\omega_{EA}^{P}=0.624$. In Fig. (a) we observe the bump in the ion $\hat{f_{i}}(v,t)$ around range of phase velocities which corresponds to QSIS inhomogeneity $k_{eq}/k_{min}=2$ and $k/k_{min}=1,3,4,5$ sideband modes listed in Table \ref{TABLE_1}. Meanwhile, in (b) local non-monotonous structures are observed in electron $\hat{f_{e}}(v,t)$ around phase velocities $v_{\phi,e,k_{p}}^{EAW}=1.560$ and $v_{\phi,e,k_{p}}^{LAN}=3.025$ which corresponds to EA and LAN modes respectively. Fig. (c) and (d) shows the zoomed variation of ion and electron $\hat{f}(v,t)$ around respective phase velocities.}
\label{1_DFE_DFI_EAW_PERT}
\end{figure}

Figs. \ref{1_ELECTRON_CP_EAW_PERT} and \ref{1_CP_ELECTRON_EAW_PERT_HOMO} illustrates variations in phase space portrait of electron distribution function $f_{e}(x,v)$ at different times for EAW perturbation driven from $120000~\omega_{pe}^{-1}<t<121000~\omega_{pe}^{-1}$ in the presence of QSIS inhomogeneity as well as homogeneous (absence of any inhomogeneity) cases with $k_{p}/k_{min}=1,~k_{min}=0.4$, $E_{0}^{P}=0.025$ and $\omega_{EA}^{P}=0.624$ respectively. In Fig. \ref{1_ELECTRON_CP_EAW_PERT} (b), around $t=121000~\omega_{pe}^{-1}$, we observe the formation of Langmuir (LAN) mode at phase velocity $v_{\phi}^{LAN}=3.21$ [since, $\omega=1.284,~k=0.4$] alongwith electron acoustic (EA) mode at phase velocity $v_{\phi}^{EA}=1.560$ [since, $\omega=0.624,~k=0.4$]. It also indicates that the formation of the LAN mode started during the EA perturbation drive between $120000~\omega_{pe}^{-1}<t<121000~\omega_{pe}^{-1}$ itself. It is obvious, the LAN and EA vortex structures in the electron phase space are robust alongwith intermediate structures in the sepratix of LAN and EA modes which can be seen in Fig. \ref{1_ELECTRON_CP_EAW_PERT} (b) and (c) around $t=121000~\omega_{pe}^{-1}$ and $123000~\omega_{pe}^{-1}$ respectively. Also, from Fig. \ref{1_ELECTRON_CP_EAW_PERT} (b) to (f), we observe the relaxation of phase space vortices (PSV) and disappearing transition of the intermediate structure till end of the simulation. It is important to note the formation and relaxation of vortex structure at velocity location $v=0$. From Fig. \ref{1_ELECTRON_CP_EAW_PERT} (f), we infer that the energy cascading process would be the prominent cause for the vanishing LAN mode and $v=0$ structure.

\begin{figure}
\centerline{\includegraphics[scale=0.70]{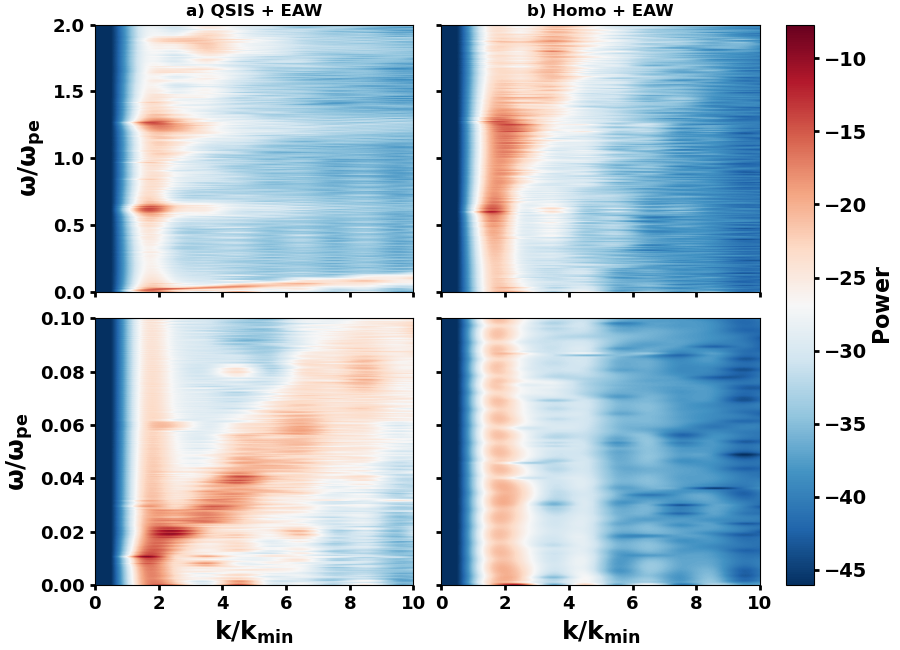}}
\caption{2D $(\omega,k)$ power spectrum plot for (a) homogeneous and (b) QSIS inhomogeneous case where EAW perturbation is applied from $0<t<121000~\omega_{pe}^{-1}$ in the presence of QSIS inhomogeneity with $k_{p}/k_{min}=1$, $E_{0}^{P}=0.025$ and $\omega_{EA}^{P}=0.624$. In (a) we observe the major power is deposited around $k/k_{min}=1$ mode and there is no significant signature of mode coupling phenomenon. From (b) we infer that the major power is deposited around QSIS inhomogeneity $k_{eq}/k_{min}$ mode and distribution of power due to mode coupling can be seen between $k_{eq}/k_{min}$, $k_{p}/k_{min}=1$ and sideband modes $k/k_{min}=3,4,5$.}
\label{1_2DPS_EAW_PERT}
\end{figure}

Meanwhile, in the homogeneous plasma case when we launched EAW mode, we have observed the formation of both LAN and EA modes respectively, as shown in Fig. \ref{1_CP_ELECTRON_EAW_PERT_HOMO}. However, the formation of the LAN mode begins after the EA drive is switched off at $t=1000~\omega_{pe}^{-1}$ as illustrated in Fig. \ref{1_CP_ELECTRON_EAW_PERT_HOMO} (b) and (c). The LAN vortex structure formed in the homogeneous case is not as prominent as it is observed in the inhomogeneous case. Alternatively, in this case, there is absence of intermediate vortex structures due to the absence of wave-wave coupling interactions which was evident in the inhomogeneous case. Also, there is absence of $v=0$ vortex structure stream in the homogeneous case contrary to the QSIS inhomogeneous case. At the end of the simulation in both homogeneous and QSIS inhomogeneous cases around times $t=10000~\omega_{pe}^{-1}$ and $t=130000~\omega_{pe}^{-1}$ respectively, we have observed some crucial differences in the final steady state such as vanishing LAN mode interior structure as well as change in the LAN mode sepratix [See Fig. \ref{1_CP_ELECTRON_EAW_PERT_HOMO} (f) and Fig. \ref{1_ELECTRON_CP_EAW_PERT} (f)]. Whereas, EA mode exists in both the cases at final times.

Fig. \ref{1_DFE_DFI_EAW_PERT} demonstrates spatially averaged distribution function $\hat{f}_{[i,e]}(v,t)$ [defined in Eq. \ref{EQ_11}] plot of (a) ions and (b) electrons at different times i.e $t=121000,~125000,~130000~\omega_{pe}^{-1}$ for EAW perturbation applied from $0<t<121000~\omega_{pe}^{-1}$ with $k_{p}/k_{min}=1$, $E_{0}^{P}=0.025$ and $\omega_{EA}^{P}=0.624$ in the presence of QSIS inhomogeneity. In Fig. \ref{1_DFE_DFI_EAW_PERT} (a) and (c) we observe the bump in the ion $\hat{f_{i}}(v,t)$ around range of phase velocities which corresponds to QSIS inhomogeneity $k_{eq}/k_{min}=2$ and $k/k_{min}=1,3,4,5$ sideband modes listed in Table \ref{TABLE_1}. In Fig. \ref{1_DFE_DFI_EAW_PERT} (b) and (d), we have observed a ``hump" or local non-monotonous structure created in the electron distribution function due to generation of EA and LAN modes whose phase velocities are around $v_{\phi,e,k_{p}}^{EAW}=1.560$ and $v_{\phi,e,k_{p}}^{LAN}=3.025$ respectively. Also, no significant changes are observed in the flat top plateau of the ion distribution function during $120~\omega_{pe}^{-1}<t<130~\omega_{pe}^{-1}$ evolution as shown in the Fig. \ref{1_DFE_DFI_EAW_PERT} (a) and (c).

Fig. \ref{1_2DPS_EAW_PERT} shows 2D $(\omega,k)$ power spectrum plot for (a) homogeneous and (b) QSIS inhomogeneous case where EAW perturbation is applied from $0<t<121000~\omega_{pe}^{-1}$ with $k_{p}/k_{min}=1$, $E_{0}^{P}=0.025$ and $\omega_{EA}^{P}=0.624$ in the presence of QSIS inhomogeneity. While comparing these two cases, from Fig. \ref{1_2DPS_EAW_PERT} (a), we have observed that major portion of the power is concentrated around $k/k_{min}=1$ mode in the homogeneous plasma case and there is a lack of any significant mode coupling signature throughout. Meanwhile, in Fig. \ref{1_2DPS_EAW_PERT} (b), for the case when EAW perturbation is launched on top of QSIS inhomogeneity, strong signature of the mode coupling phenomenon is exhibited and the fraction of the total power is distributed among several modes $k_{eq}/k_{min}=2$, $k_{p}/k_{min}=1$ and sideband modes $k/k_{min}=3,4,5$ with various frequencies.

\begin{figure}
\centerline{\includegraphics[scale=0.38]{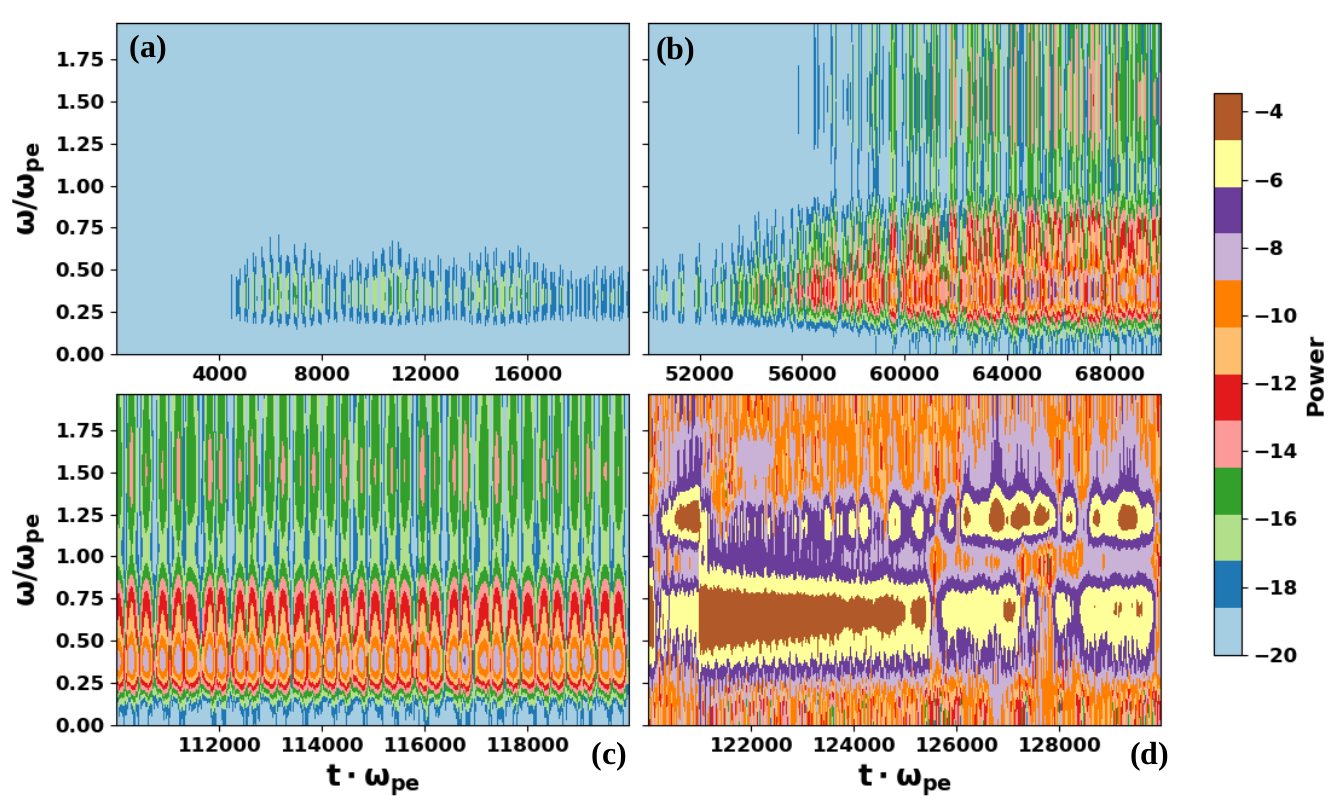}}
\caption{Portrait of spectogram i.e variation of frequency $\omega/\omega_{min}$ with respect to time for different intervals (a) $t=0$ to $20000~\omega_{pe}^{-1}$, (b) $t=50000$ to $70000~\omega_{pe}^{-1}$, (c) $t=110000$ to $120000~\omega_{pe}^{-1}$ and (d) $t=120000$ to $130000~\omega_{pe}^{-1}$ where IA drive is applied between $0<t<20000~\omega_{pe}^{-1}$, EA drive is applied between $120000~\omega_{pe}^{-1}<t<121000~\omega_{pe}^{-1}$ with $k_{p}/k_{min}=1$, $E_{0}^{D}=0.025$ and $\omega_{EA}=0.625$. From (a) to (d), we observe the generation of different frequencies during the ITPI and EAW perturbation transition.}
\label{1_SPECTOGRAM_EAW_PERT}
\end{figure}

Fig. \ref{1_SPECTOGRAM_EAW_PERT} illustrates spectogram i.e variation of frequency $\omega/\omega_{min}$ with respect to time for different intervals (a) $t=0$ to $20000~\omega_{pe}^{-1}$, (b) $t=50000$ to $70000~\omega_{pe}^{-1}$, (c) $t=110000$ to $120000~\omega_{pe}^{-1}$ and (d) $t=120000$ to $130000~\omega_{pe}^{-1}$ where IA drive is applied between $0<t<20000~\omega_{pe}^{-1}$, EA drive is applied between $120000~\omega_{pe}^{-1}<t<121000~\omega_{pe}^{-1}$ with $k_{p}/k_{min}=1$, $E_{0}^{D}=0.025$ and $\omega_{EA}=0.625$. In Fig. \ref{1_SPECTOGRAM_EAW_PERT} [(a), (b), (c) and (d)], we have observed the transition of the generated frequencies during QSIS inhomogeneity creation via ITPI process with destablization time $T_{D}^{ion}=65000~\omega_{pe}^{-1}$. Fig. \ref{1_SPECTOGRAM_EAW_PERT} (a) signifies that due to the presence of the function $g(t)$ [defined in Eq. \ref{EQ_6}], the frequency generated due to IA drive peaked around $t=4000~\omega_{pe}^{-1}$ to $16000~\omega_{pe}^{-1}$.   

As mentioned in Sec. \ref{KIKE_Equillibrium_Construction}, diverse frequency generation signature is observed around $T_{D}^{ion}=65000~\omega_{pe}^{-1}$ as shown in Fig. \ref{1_SPECTOGRAM_EAW_PERT} (c), due to the generation of the sidebands which becomes prominent to inflict ITPI at $T_{D}^{ion}=65000~\omega_{pe}^{-1}$. During relaxation period of QSIS inhomogeneous equilibrium creation, we do not observe any sudden change in the spectogram signature from $t=110000~\omega_{pe}^{-1}$ to $120000~\omega_{pe}^{-1}$ indicating smooth relaxation without any abrupt frequency response. In Fig. \ref{1_SPECTOGRAM_EAW_PERT} (d), we see the spectogram signature during the EA drive from $120000~\omega_{pe}^{-1}<t<121000~\omega_{pe}^{-1}$ till the complete evolution upto $t=130000~\omega_{pe}^{-1}$. Frequencies corresponding to EA perturbation and LAN mode i.e $\omega_{EA}^{P}=0.625$ and $\omega_{LAN}=1.21$ respectively, were generated and can be seen as two distinct bands in \ref{1_SPECTOGRAM_EAW_PERT} (d). Also, frequencies correspoding to the intermediate structures [Fig. \ref{1_ELECTRON_CP_EAW_PERT} (b)] may lie in the region between these two seprate bands of the spectogram signature.  

\begin{figure}
\centerline{\includegraphics[scale=0.65]{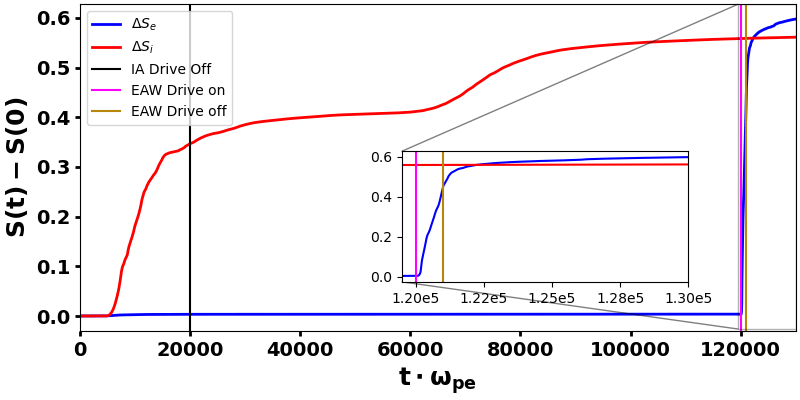}}
\caption{Variation of difference in entropy $\Delta S_{[i,e]}$ (defined in Eq. \ref{EQ_13}, \ref{EQ_14}) of ions and electrons with respect to time for EAW perturbation driven from $0<t<121000~\omega_{pe}^{-1}$ in the presence of QSIS inhomogeneity with $k_{p}/k_{min}=1$, $E_{0}^{P}=0.025$, $\omega_{EA}^{P}=0.624$ and grid sizes $[N_{x} \times N_{v}=1024 \times 6000]$. Solid line at $t=20000~\omega_{pe^{-1}}$ indicates the IA drive switch off time, at $t=120000~\omega_{pe}^{-1}$ indicates the onset of EAW drive and at $t=121000~\omega_{pe}^{-1}$ indicates the EAW drive switch off time. We observe late time saturation in the numerical entropies of both ions and electrons respectively after $t>121000~\omega_{pe}^{-1}$.}
\label{1_ENTROPY_EAW_PERT}
\end{figure}

Variation of difference in entropy $\Delta S_{[i,e]}$ (defined in Eq. \ref{EQ_13}, \ref{EQ_14}) of ions and electrons with respect to time for EAW perturbation driven from $0<t<121000~\omega_{pe}^{-1}$ in the presence of QSIS inhomogeneity with $k_{p}/k_{min}=1$, $E_{0}^{P}=0.025$, $\omega_{EA}^{P}=0.624$ and grid sizes $[N_{x} \times N_{v}=1024 \times 6000]$ is shown in Fig. \ref{1_ENTROPY_EAW_PERT} . Solid line at $t=20000~\omega_{pe}^{-1}$ indicates the IA drive switch off time, at $t=120000~\omega_{pe}^{-1}$ indicates the onset of EAW drive and at $t=121000~\omega_{pe}^{-1}$ indicates the EAW drive switch off time. We can clearly observe the increase in $\Delta S_{i}$ during Inhomogeneity creation using IA drive, meanwhile, $\Delta S_{e}$ remains close to zero when compared with $\Delta S_{i}$. During EA perturbation $120000~\omega_{pe}^{-1}<t<121000~\omega_{pe}^{-1}$, we observe a gradual increase in the $\Delta S_{e}$ signature and it attains value more than $\Delta S_{i}$ i.e $\Delta S_{e} > \Delta S_{i}$. Afterwards, $\Delta S_{e}$ saturates as shown in the inset plot signifying the chosen grid resolutions are sufficient to resolve the required physics. 

\begin{figure}
\centerline{\includegraphics[scale=0.55]{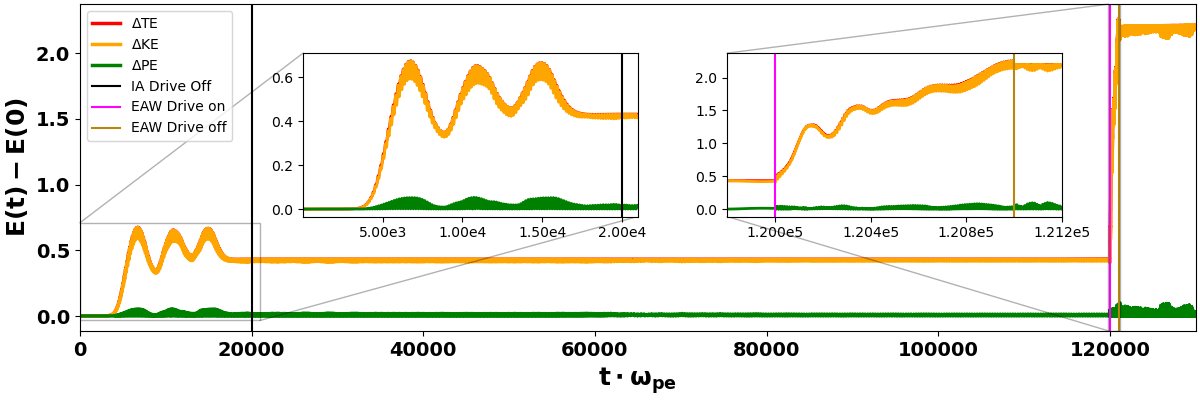}}
\caption{Relative total, kinetic and potential energies $(\Delta TE, \Delta KE_{[i,e]}, \Delta PE)$ [defined in Eq. \ref{EQ_15}, \ref{EQ_16} and \ref{EQ_17}] signatures with respect to time for the case where EAW perturbation is applied from $0<t<121000~\omega_{pe}^{-1}$ in the presence of QSIS inhomogeneity with $k_{p}/k_{min}=1$, $E_{0}^{P}=0.025$ and $\omega_{EA}^{P}=0.624$. Spatial and velocity $(x,v)$ domain grid discretization for both electrons and ions were set to $[N_{x} \times N_{v}=1024 \times 6000]$. Solid line at $t=20000~\omega_{pe}^{-1}$ indicates the IA drive switch off time, at $t=120000~\omega_{pe}^{-1}$ indicates the onset of EAW drive and at $t=121000~\omega_{pe}^{-1}$ indicates the EAW drive switch off time. Inset plot shows the zoomed relative energy variations during both the IA and EA drives respectively.}
\label{1_EC_EAW_PERT}
\end{figure}

Fig. \ref{1_EC_EAW_PERT} shows relative total, kinetic and potential energies $(\Delta TE, \Delta KE_{[i,e]}, \Delta PE)$ [defined in Eq. \ref{EQ_15}, \ref{EQ_16} and \ref{EQ_17}] signatures with respect to time for the case where EAW perturbation is applied from $0<t<121000~\omega_{pe}^{-1}$ in the presence of QSIS inhomogeneity with $k_{p}/k_{min}=1$, $E_{0}^{P}=0.025$ and $\omega_{EA}^{P}=0.624$. Spatial and velocity $(x,v)$ domain grid discretization for both electrons and ions were set to $[N_{x} \times N_{v}=1024 \times 6000]$. Solid line at $t=20000~\omega_{pe}^{-1}$ indicates the IA drive switch off time, at $t=120000~\omega_{pe}^{-1}$ indicates the onset of EAW drive and at $t=121000~\omega_{pe}^{-1}$ indicates the EAW drive switch off time. After the application of perturbative EA drive, we have observed an increase in the $(\Delta TE, \Delta KE_{[i,e]}, \Delta PE)$ signatures as shown in Fig. \ref{1_EC_EAW_PERT}. Inset plot shows the huge rise in $\Delta KE_{[i,e]}$ which leads to the rise in $\Delta TE $ eventually due to EA perturbation. However, at late times the energy signatures attains a constant value indicating good energy conservation for the obtained steady state solution. It also signifies that the grid resolution $[N_{x} \times N_{v}]$ is sufficent enough to attain the conservation.


\section{Discussion and Conclusion}
\label{KIKE_Discussion_conclusion}

In this work i.e Part-I, we have investigated the response of electron acoustic wave (EAW) launched in the presence of QSIS inhomogeneity created by an external electric field drive at IA frequency such that the electron remain Maxwellian throughout the drive period. Using VPPM-OMP 1.0 Vlasov-Poisson solver with kinetic ions and kinetic electrons, we have created an equilibrium QSIS inhomogeneity with IA drive defined in Eq. \ref{EQ_5}. Analogous to the large amplitude electron plasma waves, during the process of formation of quasi-steady ion scale inhomogeneous state, we have observed instability due to trapped ions in the ion phase space (PS) and termed it as ion trapped particle instability (ITPI). Meanwhile, unperturbed electron Maxwellian distribution were maintained due to the close to the Maxwellian distribution nature of the IA drive. 

Generation of sideband modes which are coupled to the QSIS scale length i.e $k_{eq}$ occurs due to non-linear perturbation amplitude, leading to the energy exchange between interacting modes via wave-wave mode coupling phenomenon. This leads to the amplitude equivalence between $k_{eq}$ and sideband modes which initiates the destabilization process around ion destabilization time $T_{D}^{ion}=65000~\omega_{pe}^{-1}$. As a result, we have observed the transition from m=2 to m=1 mode due to energy cascading. Extended simulations were performed upto $t=120000~\omega_{pe}^{-1}$ to obtain a QSIS inhomogeneity. A very unique feature of these long time QSIS inhomogeneity creation simulations are that the IA drive is applied in such a way which does not perturb the electron velocity distribution throughout from $t=0$ to $t=120000~\omega_{pe}^{-1}$ as shown in the Fig. \ref{1_ELECTRON_CP_EQC}. Energy conservation and entropy diagnostics suggested that obtained QSIS inhomogeneity profile is a steady state with saturated electric field value of 0.04.

Using the exact simulation parameters, we have noticed following interesting differences between two cases, i.e when the EA wave is launched in the homogeneous ion spatial profile and inhomogeneous ion spatial profile plasmas respectively,
\begin{itemize}
	\item In inhomogeneous case, Langmuir mode was generated in the electron phase space during EA driving interval $120000~\omega_{pe}^{-1}<t<121000~\omega_{pe}^{-1}$ which was absent in the homogeneous case.
	
	\item Upon addition of EA perturbation, generation of intermediate vortex structures occured in the QSIS inhomogeneous case, as against the homogeneous case. Wave-wave interaction due to the presence of the QSIS inhomogeneity can be the probable reason behind this observation.
	
	\item In case of inhomogeneous plasma, we have also observed the formation of transient vortex structure at $v=0$ in the electron phase space during the relaxation period. But there was no such occurrence for homogeneous case as shown in Fig. \ref{1_ELECTRON_CP_EAW_PERT} and Fig. \ref{1_CP_ELECTRON_EAW_PERT_HOMO} respectively.
	
	\item From 2D $(\omega,k)$ power spectrum plot i.e Fig. \ref{1_2DPS_EAW_PERT}, we inferred the absence and presence of wave-wave mode coupling phenomenon in either homogeneous or inhomogeneous ion spatial profile cases.
	
	\item In addition, spectrogram [Fig. \ref{1_SPECTOGRAM_EAW_PERT}] of QSIS inhomogeneous case suggests various frequency generation throughout, which supports the previous observations. Separate EA and LAN frequency bands were also forms as shown in Fig. \ref{1_SPECTOGRAM_EAW_PERT}.
\end{itemize} 

In the companion paper i.e Part-II, we will present our investigations for the dynamics of large phase space vortices (PSV) driven with time dependent $\omega(t)$ or chriped frequency in the presence of the QSIS inhomogeneity created in this Part-I. Also, we will be highlighting the key comparative differences of these chriped driven vortex structures in the absence of any ion scale inhomogeneity.


\section*{Acknowledgments}
All the computational results of this paper were obtained using the ANTYA HPC Linux cluster at Institute for Plasma Research (IPR) Gandhinagar, India. The authors would like to thank the Data Center staff at IPR.


\section*{Data availability statement}
The data that support the findings of this study are available upon reasonable request from the authors.

\section*{References}
\bibliography{iopart-num}

\providecommand{\newblock}{}
\begin{thebibliography}{10}
\expandafter\ifx\csname url\endcsname\relax
  \def\url#1{{\tt #1}}\fi
\expandafter\ifx\csname urlprefix\endcsname\relax\def\urlprefix{URL }\fi
\providecommand{\eprint}[2][]{\url{#2}}

\bibitem{bgk1957}
Bernstein I~B, Greene J~M and Kruskal M~D 1957 {\em Phys. Rev.\/} {\bf 108}(3)
  546--550 \urlprefix\url{https://link.aps.org/doi/10.1103/PhysRev.108.546}

\bibitem{landau1946}
Haar D 2013 {\em Collected Papers of L.D. Landau\/} (Elsevier Science) ISBN
  9781483152707
  \urlprefix\url{https://books.google.co.in/books?id=epc4BQAAQBAJ}

\bibitem{oneil1965}
O'Neil T 1965 {\em The Physics of Fluids\/} {\bf 8} 2255--2262
  (\textit{Preprint}
  \eprint{https://aip.scitation.org/doi/pdf/10.1063/1.1761193})
  \urlprefix\url{https://aip.scitation.org/doi/abs/10.1063/1.1761193}

\bibitem{kds1969}
Kruer W~L, Dawson J~M and Sudan R~N 1969 {\em Phys. Rev. Lett.\/} {\bf 23}(15)
  838--841 \urlprefix\url{https://link.aps.org/doi/10.1103/PhysRevLett.23.838}

\bibitem{kd1970}
Kruer W~L and Dawson J~M 1970 {\em The Physics of Fluids\/} {\bf 13} 2747--2751
  (\textit{Preprint}
  \eprint{https://aip.scitation.org/doi/pdf/10.1063/1.1692859})
  \urlprefix\url{https://aip.scitation.org/doi/abs/10.1063/1.1692859}

\bibitem{goldman1970}
Goldman M~V 1970 {\em The Physics of Fluids\/} {\bf 13} 1281--1289
  (\textit{Preprint}
  \eprint{https://aip.scitation.org/doi/pdf/10.1063/1.1693061})
  \urlprefix\url{https://aip.scitation.org/doi/abs/10.1063/1.1693061}

\bibitem{rosen1972}
Rosen B, Schmidt G and Kruer W~L 1972 {\em The Physics of Fluids\/} {\bf 15}
  2001--2006 (\textit{Preprint}
  \eprint{https://aip.scitation.org/doi/pdf/10.1063/1.1693814})
  \urlprefix\url{https://aip.scitation.org/doi/abs/10.1063/1.1693814}

\bibitem{Schamel_1975}
Schamel H 1975 {\em Journal of Plasma Physics\/} {\bf 13} 139–145

\bibitem{canosa1976}
Canosa J and Wray A 1976 {\em The Physics of Fluids\/} {\bf 19} 1958--1966
  (\textit{Preprint}
  \eprint{https://aip.scitation.org/doi/pdf/10.1063/1.861413})
  \urlprefix\url{https://aip.scitation.org/doi/abs/10.1063/1.861413}

\bibitem{shoucri1978}
Shoucri M~M 1978 {\em The Physics of Fluids\/} {\bf 21} 1359--1365
  (\textit{Preprint}
  \eprint{https://aip.scitation.org/doi/pdf/10.1063/1.862377})
  \urlprefix\url{https://aip.scitation.org/doi/abs/10.1063/1.862377}

\bibitem{shoucri1980}
Shoucri M 1980 {\em The Physics of Fluids\/} {\bf 23} 2030--2033
  (\textit{Preprint}
  \eprint{https://aip.scitation.org/doi/pdf/10.1063/1.862889})
  \urlprefix\url{https://aip.scitation.org/doi/abs/10.1063/1.862889}

\bibitem{schamel1982}
Schamel H 1982 {\em Phys. Rev. Lett.\/} {\bf 48}(7) 481--483
  \urlprefix\url{https://link.aps.org/doi/10.1103/PhysRevLett.48.481}

\bibitem{koch1983}
Koch B~P and Leven R~W 1983 {\em Physica Scripta\/} {\bf 27} 220--224
  \urlprefix\url{https://doi.org/10.1088/0031-8949/27/3/013}

\bibitem{ghizzo1988}
Ghizzo A, Izrar B, Bertrand P, Fijalkow E, Feix M~R and Shoucri M 1988 {\em The
  Physics of Fluids\/} {\bf 31} 72--82 (\textit{Preprint}
  \eprint{https://aip.scitation.org/doi/pdf/10.1063/1.866579})
  \urlprefix\url{https://aip.scitation.org/doi/abs/10.1063/1.866579}

\bibitem{manfredi1997}
Manfredi G 1997 {\em Phys. Rev. Lett.\/} {\bf 79}(15) 2815--2818
  \urlprefix\url{https://link.aps.org/doi/10.1103/PhysRevLett.79.2815}

\bibitem{manfredi2000}
Manfredi G and Bertrand P 2000 {\em Physics of Plasmas\/} {\bf 7} 2425--2431
  (\textit{Preprint} \eprint{https://doi.org/10.1063/1.874081})
  \urlprefix\url{https://doi.org/10.1063/1.874081}

\bibitem{brunetti2000}
Brunetti M, Califano F and Pegoraro F 2000 {\em Phys. Rev. E\/} {\bf 62}(3)
  4109--4114 \urlprefix\url{https://link.aps.org/doi/10.1103/PhysRevE.62.4109}

\bibitem{brunner2004}
Brunner S and Valeo E~J 2004 {\em Phys. Rev. Lett.\/} {\bf 93}(14) 145003
  \urlprefix\url{https://link.aps.org/doi/10.1103/PhysRevLett.93.145003}

\bibitem{shoucri_2006}
Shoucri M 2006 {\em Journal of Plasma Physics\/} {\bf 72} 861–864

\bibitem{brunner2014}
Brunner S, Berger R~L, Cohen B~I, Hausammann L and Valeo E~J 2014 {\em Physics
  of Plasmas\/} {\bf 21} 102104 (\textit{Preprint}
  \eprint{https://doi.org/10.1063/1.4896753})
  \urlprefix\url{https://doi.org/10.1063/1.4896753}

\bibitem{shoucri2017}
Shoucri M 2017 {\em Laser and Particle Beams\/} {\bf 35} 706–721

\bibitem{yang2020}
Yang T, Feng Q~S, Wang Y~X, Zhou Y~Z, Ban S~S, Zhang S~T, Xie R, Jiang Y, Cao
  L~H, Liu Z~J and Zheng C~Y 2020 {\em Plasma Physics and Controlled Fusion\/}
  {\bf 62} 095009 \urlprefix\url{https://doi.org/10.1088/1361-6587/ab9d68}

\bibitem{Pandey_2021_TPI_1}
Pandey S~K and Ganesh R 2021 {\em Physica Scripta\/} {\bf 96} 125616
  \urlprefix\url{https://doi.org/10.1088/1402-4896/ac25a1}

\bibitem{Pandey_2021_TPI_2}
Pandey S~K and Ganesh R 2021 {\em Physica Scripta\/} {\bf 96} 125615
  \urlprefix\url{https://doi.org/10.1088/1402-4896/ac25a2}

\bibitem{Temerin_1982}
Temerin M, Cerny K, Lotko W and Mozer F~S 1982 {\em Phys. Rev. Lett.\/} {\bf
  48}(17) 1175--1179
  \urlprefix\url{https://link.aps.org/doi/10.1103/PhysRevLett.48.1175}

\bibitem{Franz_1998}
Franz J~R, Kintner P~M and Pickett J~S 1998 {\em Geophysical Research
  Letters\/} {\bf 25} 1277--1280 (\textit{Preprint}
  \eprint{https://agupubs.onlinelibrary.wiley.com/doi/pdf/10.1029/98GL50870})
  \urlprefix\url{https://agupubs.onlinelibrary.wiley.com/doi/abs/10.1029/98GL50870}

\bibitem{Mangeney_1999}
Mangeney A, Salem C, Lacombe C, Bougeret J~L, Perche C, Manning R, Kellogg P~J,
  Goetz K, Monson S~J and Bosqued J~M 1999 {\em Annales Geophysicae\/} {\bf 17}
  307--320 \urlprefix\url{https://angeo.copernicus.org/articles/17/307/1999/}

\bibitem{lynov_1979}
Lynov J~P, Michelsen P, Pécseli H~L, Rasmussen J~J, Saéki K and Turikov V~A
  1979 {\em Physica Scripta\/} {\bf 20} 328
  \urlprefix\url{https://doi.org/10.1088/0031-8949/20/3-4/005}

\bibitem{Saeki_1979}
Saeki K, Michelsen P, P\'ecseli H~L and Rasmussen J~J 1979 {\em Phys. Rev.
  Lett.\/} {\bf 42}(8) 501--504
  \urlprefix\url{https://link.aps.org/doi/10.1103/PhysRevLett.42.501}

\bibitem{Danielson_2004}
Danielson J~R, Anderegg F and Driscoll C~F 2004 {\em Phys. Rev. Lett.\/} {\bf
  92}(24) 245003
  \urlprefix\url{https://link.aps.org/doi/10.1103/PhysRevLett.92.245003}

\bibitem{raghunathan2013}
Raghunathan M and Ganesh R 2013 {\em Physics of Plasmas\/} {\bf 20} 032106
  (\textit{Preprint} \eprint{https://doi.org/10.1063/1.4794320})
  \urlprefix\url{https://doi.org/10.1063/1.4794320}

\bibitem{pallavithesis}
Trivedi P 2019 {\em Driven Phase Space Structures In A 1D Vlasov-Poisson
  Plasma\/} Ph.D. thesis Institute for Plasma Research

\bibitem{Holloway_Dorning_1991}
Holloway J~P and Dorning J~J 1991 {\em Phys. Rev. A\/} {\bf 44}(6) 3856--3868
  \urlprefix\url{https://link.aps.org/doi/10.1103/PhysRevA.44.3856}

\bibitem{valentini_2006}
Valentini F, O’Neil T~M and Dubin D~H~E 2006 {\em Physics of Plasmas\/} {\bf
  13} 052303 ISSN 1070-664X \urlprefix\url{https://doi.org/10.1063/1.2198467}

\bibitem{valentni_2025}
Valentini F, O'Neil T~M and Dubin D~H 2025 {\em Physics of Plasmas\/} {\bf 32}
  042104 ISSN 1070-664X \urlprefix\url{https://doi.org/10.1063/5.0256797}

\bibitem{Anderegg_2009}
Anderegg F, Driscoll C~F, Dubin D~H~E, O’Neil T~M and Valentini F 2009 {\em
  Physics of Plasmas\/} {\bf 16} 055705 ISSN 1070-664X
  \urlprefix\url{https://doi.org/10.1063/1.3099646}

\bibitem{Anderegg_PRL_2009}
Anderegg F, Driscoll C~F, Dubin D~H~E and O'Neil T~M 2009 {\em Phys. Rev.
  Lett.\/} {\bf 102}(9) 095001
  \urlprefix\url{https://link.aps.org/doi/10.1103/PhysRevLett.102.095001}

\bibitem{Rivera_2025}
Rivera D~M, Carril H~A, Araneda J~A and Navarro R~E 2025 {\em Physica
  Scripta\/} {\bf 100} 105605
  \urlprefix\url{https://doi.org/10.1088/1402-4896/ae0c47}

\bibitem{Breizman_1997}
Breizman B~N, Berk H~L, Pekker M~S, Porcelli F, Stupakov G~V and Wong K~L 1997
  {\em Physics of Plasmas\/} {\bf 4} 1559--1568 ISSN 1070-664X
  \urlprefix\url{https://doi.org/10.1063/1.872286}

\bibitem{Eremin_2002}
Eremin D~Y and Berk H~L 2002 {\em Physics of Plasmas\/} {\bf 9} 772--785 ISSN
  1070-664X \urlprefix\url{https://doi.org/10.1063/1.1436492}

\bibitem{Fajans_2003}
Bertsche W, Fajans J and Friedland L 2003 {\em Phys. Rev. Lett.\/} {\bf 91}(26)
  265003 \urlprefix\url{https://link.aps.org/doi/10.1103/PhysRevLett.91.265003}

\bibitem{Friedland_2004}
Friedland L, Peinetti F, Bertsche W, Fajans J and Wurtele J 2004 {\em Physics
  of Plasmas\/} {\bf 11} 4305--4317 ISSN 1070-664X
  \urlprefix\url{https://doi.org/10.1063/1.1781166}

\bibitem{Peinetti_2005}
Peinetti F, Bertsche W, Fajans J, Wurtele J and Friedland L 2005 {\em Physics
  of Plasmas\/} {\bf 12} 062112 ISSN 1070-664X
  \urlprefix\url{https://doi.org/10.1063/1.1928251}

\bibitem{pallavi2016}
Trivedi P and Ganesh R 2016 {\em Physics of Plasmas\/} {\bf 23} 062112
  (\textit{Preprint} \eprint{https://doi.org/10.1063/1.4953603})
  \urlprefix\url{https://doi.org/10.1063/1.4953603}

\bibitem{pallavi2017}
Trivedi P and Ganesh R 2017 {\em Physics of Plasmas\/} {\bf 24} 032107
  (\textit{Preprint} \eprint{https://doi.org/10.1063/1.4978560})
  \urlprefix\url{https://doi.org/10.1063/1.4978560}

\bibitem{sanjeev2021}
Pandey S~K and Ganesh R 2021 {\em AIP Advances\/} {\bf 11} 025229
  (\textit{Preprint} \eprint{https://doi.org/10.1063/5.0030082})
  \urlprefix\url{https://doi.org/10.1063/5.0030082}

\bibitem{pandey_2022_KAW}
Pandey S~K, Mahapatra J and Ganesh R 2022 {\em Physica Scripta\/} {\bf 97}
  105602 \urlprefix\url{https://dx.doi.org/10.1088/1402-4896/ac90f4}

\bibitem{Pandey_2024}
Pandey S~K and Ganesh R 2024 {\em Physica Scripta\/} {\bf 99} 125608
  \urlprefix\url{https://doi.org/10.1088/1402-4896/ad8cae}

\bibitem{Saini2018}
Saini V, Pandey S~K, Trivedi P and Ganesh R 2018 {\em Physics of Plasmas\/}
  {\bf 25} 092107 (\textit{Preprint}
  \eprint{https://doi.org/10.1063/1.5024376})
  \urlprefix\url{https://doi.org/10.1063/1.5024376}

\bibitem{sanjeevthesis}
Pandey S~K 2023 {\em Linear and non-linear waves in spatially non-uniform 1D
  Vlasov-Poisson plasmas.\/} Ph.D. thesis Institute for Plasma Research

\bibitem{colella1984}
Colella P and Woodward P~R 1984 {\em Journal of Computational Physics\/} {\bf
  54} 174 -- 201 ISSN 0021-9991
  \urlprefix\url{http://www.sciencedirect.com/science/article/pii/0021999184901438}

\bibitem{cheng1976}
Cheng C and Knorr G 1976 {\em Journal of Computational Physics\/} {\bf 22} 330
  -- 351 ISSN 0021-9991
  \urlprefix\url{http://www.sciencedirect.com/science/article/pii/002199917690053X}

\bibitem{Chen}
Chen F~F 2018 {\em Introduction to Plasma Physics and Controlled Fusion\/}
  (Springer Cham Heidelberg New York Dordrecht London)

\bibitem{feix}
Feix M~R, Bertrand P and Ghizzo A 1994 {\em Advances in Kinetic Theory and
  Computing\/} (World Scientific, Singapore. pp. 45–81.)

\bibitem{kaw1973}
Kaw P~K, Lin A~T and Dawson J~M 1973 {\em The Physics of Fluids\/} {\bf 16}
  1967--1975 (\textit{Preprint}
  \eprint{https://aip.scitation.org/doi/pdf/10.1063/1.1694242})
  \urlprefix\url{https://aip.scitation.org/doi/abs/10.1063/1.1694242}

\end{thebibliography}

\end{document}